\documentclass[pra,twocolumn,showpacs,amsmath,amssymb,superscriptaddress,notitlepage]{revtex4-1}
\usepackage{psfrag,graphicx}
\usepackage{amsfonts,amssymb,amsmath}      
\usepackage{graphicx}
\usepackage{dcolumn}
\usepackage{bm}
\usepackage{float}
\usepackage{hyperref}
\usepackage{accents}
\usepackage{bbding}
\usepackage{color,soul}
\usepackage{epstopdf}

\newcommand{\ellipse}{\raisebox{-1pt}{\scalebox{1.3}[.4]{$\circ$}}}
\newcommand{\halo}{\accentset{\ellipse}}
\newcommand{\erf}[1]{Eq.~(\ref{#1})}
\newcommand{\beq}{\begin{equation}}
\newcommand{\eeq}{\end{equation}}

\newcommand{\erfs}[2]{Eqs.~(\ref{#1})--(\ref{#2})}

\newcommand{\dg}{^\dagger}

\newcommand{\sch}{Schr\"odinger}

\newcommand{\Tr}{\text{Tr}}

\newcommand{\tp}{^{\top}}

\renewcommand{\c}{_{\text{c}}}
\newcommand{\ob}{_{\text{o}}}
\newcommand{\un}{_{\text{u}}}
\newcommand{\m}{_{\text{m}}}
\newcommand{\p}{_{\text{p}}}
\newcommand{\ex}[1]{\langle{#1}\rangle}
\newcommand{\dd}{{\rm d}}

\newcommand{\ddt}[1]{\frac{\dd{#1}}{\dd t}}

\newcommand{\SHUR}{\sch-Heisenberg uncertainty relation}

\newcommand{\past}[1]{\overleftarrow{#1}}%\overleftarrow{#1}}
\newcommand{\fut}[1]{\overrightarrow{#1}}
\newcommand{\both}[1]{\overleftrightarrow{#1}}
\newcommand{\fil}{_{\text F}}
\newcommand{\rfil}{_{\text R}}
\newcommand{\sm}{_{\text S}}

\newcommand{\god}{_{\text T}}%{^{\rm true}}%{_{\text G}}
\newcommand{\inv}{^{-1}}
\newcommand{\bx}{{\bf x}}
\newcommand{\bz}{{\bf z}}
\newcommand{\bcx}{{\check{ \bf x}}}
\newcommand{\by}{{\bf y}}
\newcommand{\bv}{{\bf v}}
\newcommand{\bw}{{\bf w}}

\newcommand{\hV}{\halo{V}}

\newcommand{\xfil}{\ex{\hat\bx}\fil}

\newcommand{\xsm}{\ex{\hat\bx}\sm}
\newcommand{\zsm}{\ex{\hat\bz}\sm}
\newcommand{\xgod}{\ex{\hat\bx}\god}
\newcommand{\K}{{\cal K}}

\newcommand{\rr}{_{{\rm r}}}
\newcommand{\cxf}{\ex{\bx}\fil}

\newcommand{\cxs}{\ex{\bx}\sm}

\definecolor{nblue}{rgb}{0.06,0.3,0.73}%229 11R, 61G, 145B
\definecolor{nblack}{rgb}{0,0,0}
\definecolor{nred}{rgb}{0.9,0.1,0.1}
\definecolor{nmagenta}{rgb}{0.7,0.0,0.3}

\newcommand{\hbx}{\hat\bx}

\begin{document}
\title{The Quantum Rauch-Tung-Striebel Smoothed State}
\author{Kiarn T. Laverick}
\affiliation{Centre for Quantum Computation and Communication Technology 
(Australian Research Council), Centre for Quantum Dynamics, Griffith University, Nathan, Queensland 
4111, Australia (email: k.laverick@griffith.edu.au)}

\begin{abstract}
Smoothing is a technique that estimates the state of a system using measurement 
information both prior and posterior to the estimation time. Two notable examples of this technique are the 
Rauch-Tung-Striebel and Mayne-Fraser-Potter smoothing techniques for linear Gaussian systems, both 
resulting in the optimal smoothed estimate of the state. However, when considering a quantum system, 
classical smoothing techniques can result in an estimate that is not a valid quantum state. Consequently, a 
different smoothing theory was developed explicitly for quantum systems. This theory has since been applied to 
the special case of linear Gaussian quantum (LGQ) systems, where, in deriving the LGQ state smoothing 
equations, the Mayne-Fraser-Potter technique was utilised. As a result, the final equations describing 
the smoothed state are closely related to the classical Mayne-Fraser-Potter smoothing equations. In this 
paper, I derive the equivalent Rauch-Tung-Striebel form of the quantum state smoothing equations, which 
further simplify the calculation for the smoothed quantum state in LGQ systems. 
Additionally, the new form of the LGQ smoothing equations bring to light a property of the smoothed quantum 
state that was hidden in the Mayne-Fraser-Potter form, the non-differentiablilty of the smoothed mean. By 
identifying the non-differentiable part of the smoothed mean, I was then able to derive a necessary and 
sufficient condition for the quantum smoothed mean to be differentiable in the steady state regime.
\end{abstract}
\maketitle

\section{Introduction}
Estimating an unknown state, i.e., a probability density function (PDF),
of physical systems using indirect measurement results has been studied 
in great depth 
\cite{KalBuc61, Rauch63, RTS65, Mayne66, Fraser67, FraPot69, Weinert01, Hay01, vanTrees13, BroHwa12, Einicke12, Fri12, Sarkka13}. 
When restricting to the case of (classical) 
linear Gaussian (LG) systems, Kalman and Bucy \cite{KalBuc61}
developed an optimal estimation technique, known as {\em filtering}, which conditions the estimate of the 
state on past measurement information, i.e., measurement information up until the time of estimation $\tau$. 
This estimated state is referred to as the filtered state. While one of the appeals of the Kalman-Bucy filtering 
technique is its ability to actively update the estimate of the state in real-time, there are other optimal estimation 
techniques that are more accurate 
\cite{Rauch63, RTS65, Mayne66, Fraser67, FraPot69, Weinert01, Hay01, vanTrees13, BroHwa12, Einicke12, Fri12, Sarkka13}. 

One such technique was developed soon after the Kalman-Bucy filtering theory by Rauch, Tung and 
Striebel \cite{Rauch63, RTS65}. 
This technique, referred to as {\em smoothing}, not only utilises the past measurement information as the 
Kalman-Bucy filter does, but also uses information gathered after the estimation time $\tau$, i.e. the `future' 
measurement record, to provide a more accurate estimate of the state. The Rauch-Tung-Striebel (RTS) 
smoothing technique, first, uses the Kalman-Bucy filtering technique to estimate the 
state until the final estimation time $T$. Once at the final estimation time $T$, the RTS smoothing equations 
run back over the estimated state, updating the results based on the future measurement record that has been 
gathered. This results in a smoothed estimate of the state. The only drawback of this smoothing technique, 
compared to filtering, is that the smoothed state cannot be obtained in real-time since future information is 
required.

A similar technique was developed by Mayne \cite{Mayne66}, Fraser \cite{Fraser67} and Potter \cite{FraPot69} 
shortly after RTS smoothing which also utilised a past-future measurement record. Mayne-Fraser-Potter (MFP) 
smoothing, sometimes referred to as two-filter smoothing 
for reasons that will become apparent, also utilises Kalman-Bucy filtering to condition on the past measurement 
record. However, to make use of the future information, they introduced a secondary filter, which I will refer to 
as a {\em retrofilter}, that ran backwards from a final uninformative state conditioning on the 
the future measurement record. This state is the retrofiltered state. Combining the filtered state and retrofiltered 
state together results in the MFP smoothed state. It has since been shown \cite{FraPot69} that the RTS 
smoothed state and the MFP smoothed state are, in fact, identical.

Moving to quantum systems, an analogous problem to classical state estimation also exists, where instead of 
estimating a PDF one wishes to estimate the density matrix $\rho$. Similar to the classical filtering theory, 
one can estimate the quantum state based on the past measurement record. As a result, this technique is often 
referred to as quantum filtering \cite{Bel87,Bel92,Bel99}. 
Interestingly, if one restricts to linear Gaussian quantum (LGQ) 
systems, the quantum filtering technique reduces to the Kalman-Bucy filtering theory 
\cite{DohJac99,WisDoh05,WisMil10}. 
Given this, one might assume that classical smoothing techniques, like the 
RTS or MFP techniques, can also be applied to quantum state estimation. This is not 
the case. Applying the classical theory to quantum state estimation can result in an unphysical estimate of the 
quantum state \cite{Tsang-PRA09,GJM13,GueWis15,Ohki15,LCW-QS20}. 
This is due to the 
non-commutativity of the operators describing the system and the operators describing the future measurement 
outcomes in quantum theory, which is usually not present in the 
classical theory. Due to the failure of the classical theory, Guevara and Wiseman \cite{GueWis15} devised a 
new smoothing theory specifically for quantum systems, the quantum state smoothing theory. 

The quantum state smoothing theory \cite{GueWis15,CGW19,LCW-QS20} introduces a secondary 
measurement record that is unobserved by the observer, say Alice, but is observed by someone else, say Bob. 
The role of Bob's measurement is to gather information about the system that Alice's measurement may have 
missed. If Alice had access to Bob's record, 
hereby referred to as the `unobserved' record, she 
could condition the estimate of the state on both her past observed record $\past{\bf O}$ and unobserved 
record $\past{\bf U}$ to obtain the true state of the system $\rho\god = \rho_{\past{\bf O}\past{\bf U}}$, a state 
containing the maximum amount of information about the system given the measurements. Here, the leftward 
arrow indicates that it is the past record and a bidirectional arrow will indicate a past-future record. However, 
since Alice does not have access to the unobserved record she cannot compute the true state of the system. 
The best \cite{CGLW21} Alice can do is to estimate the true state based only on the observed record. Now, Alice 
can obtain a smoothed quantum state by averaging over all possible true states conditioned on her past-future 
observed record $\both{\bf O}$ 
\cite{GueWis15}, i.e.,
\beq
\rho\sm = \mathbb{E}_{\past{\bf U}|\both{\bf O}}\{\rho\god\}\,, \label{qsm}
\eeq
where $\mathbb{E}_{A|B}\{C\}$ denotes the ensemble average of $C$ over $A$ conditioned on $B$, and may 
appear without the $A$ subscript when $A = C$.

Following the conception of the smoothed quantum state, the theory was applied to the special case 
of LGQ systems \cite{LCW19,LCW-QS20,LCW-PRA21}. In order to derive the LGQ smoothed quantum state, 
the classical MFP smoothing techniques was used. These quantum state smoothing equations, due to being 
closed-form equations, have been able to identify properties of the smoothed quantum state 
\cite{LCW19,LCW-QS20,LCW-PRA21} that would have been difficult to find in the general case. Furthermore, 
the smoothed state is far simpler to compute for LGQ systems compared to even a simple system in the 
general case, allowing for easier verification of these properties. However, while the MFP forms of the quantum 
state smoothing equations have been very useful, they require the calculation of a retrofiltered state. This can 
be avoided by instead using the RTS form of the smoothing equations, making the smoothed 
state even simpler to compute for LGQ systems. Additionally, the RTS form of the quantum state smoothing 
equations are dynamical equation and can provide insight into properties of the smoothed quantum state that 
would otherwise be hidden in the MFP form. Case in point, I derive a necessary and sufficient constraint for 
mean of the smoothed quantum state evolve smoothly in the steady state regime.

This paper is structured as follows. First, in Sec.~\ref{Sec-LG}, I will briefly review classical and quantum state 
smoothing for linear Gaussian systems. Next, in Sec.~\ref{Sec-Derivation} will derive the RTS form of the 
smoothed quantum state. Finally, in Sec.~\ref{Sec-Disc}, I will discuss the new form of the LGQ state smoothing 
equations, showing that, under the same condition that makes the classically smoothed mean a continuous 
function, the path of the smoothed mean is necessarily non-differentiable. Furthermore, I identify a necessary and 
sufficient condition for the mean of the smoothed quantum state to be differentiable in the steady state regime.

\section{Linear Gaussian State Estimation}\label{Sec-LG}
\subsection{Classical}
Consider a classical dynamical system. The state of the system is given by a probability density function 
$\wp(\bcx)$, where $\bcx = (\check x_1,\check x_2,...,\check x_M)\tp$ is a vector of $M$ parameters that are 
required to completely characterize the system and $\top$ denotes the transpose. Note, for clarity, the wedge 
accent will be used to denote a dummy variable to differentiate it from the corresponding random variable. For a LG system, the 
classical state will be a Gaussian distribution, i.e., $\wp(\bcx) = g(\bcx;\mathbb{E}\{\bcx\},V)$, completely 
described by its mean $\mathbb{E}\{\bx\}$ and covariance matrix 
$V = \mathbb{E}\{\bcx\bcx\tp\} - \mathbb{E}\{\bcx\}\mathbb{E}\{\bcx\tp\}$. To 
guarantee that the state remains Gaussian throughout the evolution of the 
system, provided that at the initial time $t_0$ the state is Gaussian with mean $\mathbb{E}\{\bx(t_0)\} = \bx_0$ 
and covariance $V(t_0) = V_0$, it is necessary that the following two constraints are satisfied 
\cite{Weinert01,Hay01,vanTrees13,BroHwa12,Einicke12,Fri12,WisMil10}. Firstly, the evolution of $\bx$ must be 
described by a linear Langevin equation
\beq
\dd\bx = A\bx\dd t + E\dd\bv\p\,. \label{LLE}
\eeq
Here $A$ (the drift matrix) and $E$ are constant matrices and the process noise $\dd\bv\p$ is a 
vector of independent Weiner increments satisfying 
\beq\label{Noise_cond}
\mathbb{E}\{\dd\bv\p\} = 0\,,\qquad \dd\bv\p\dd\bv\p\tp = I_{k}\dd t\,,
\eeq
where $I_k$ 
is the $k\times k$ identity matrix. Secondly, any measurement current $\by$ used to refine our estimate 
of the state must also be linear, i.e.,
\beq\label{c_meas}
\by\dd t = C\bx\dd t + \dd\bv\m\,,
\eeq
where the measurement matrix $C$ is a constant matrix and the measurement noise $\dd\bv\m$ is a 
vector of independent Wiener increments satisfying similar conditions to \erf{Noise_cond}. Usually, it is also
assumed that the process noise and measurement noise are uncorrelated 
\cite{KalBuc61,Rauch63,RTS65,Mayne66,Fraser67,FraPot69,Weinert01,Hay01,BroHwa12,Einicke12,Fri12,vanTrees13}, 
i.e., $\dd\bv\p(\dd\bv\m)\tp = 0$, however, this assumption is not always true, i.e., there might be measurement 
backaction. For the sake of generality I will consider the case where the two noises may be correlated, 
described by the cross-correlation matrix \cite{KaiFro68,Kai70,Kai73,BLP79,WisMil10} 
\beq
\Gamma\tp \dd t = E\dd\bv\p(\dd\bv\m)\tp.
\eeq 

If all of the above criteria are satisfied, one can calculate the filtered estimate of the state 
$\wp\fil(\bcx) \equiv \wp(\bcx|\past{\bf O}) = g(\bcx;\bx\fil,V\fil)$ by conditioning the state on the past 
measurement record $\past{\bf O}$. The filtered mean $\bx\fil := \mathbb{E}_{|\past{\bf O}}\{\bcx\}$ and 
covariance $V\fil = \mathbb{E}_{\bcx|\past{\bf O}}\{\bcx\bcx\tp\} - \bx\fil\bx\fil\tp$ are given by the Kalman-Bucy filtering 
equations \cite{KalBuc61,WisMil10,KaiFro68,Kai70,Kai73,BLP79}
\begin{align}
&\dd\bx\fil = A\bx\fil\dd t + \K^+[V\fil]\dd\bw\fil\,,\label{xfil}\\
&\ddt{V\fil} = AV\fil + V\fil A\tp + D - \K^+[V\fil]\K^+[V\fil]\tp\,, \label{Vfil}
\end{align}
with initial conditions $\bx\fil(t_0) = \bx_0$ and $V\fil(t_0) = V_0$.
Here $D = EE\tp$ is the diffusion matrix,
\beq\label{gain}
\K^\pm[V] = V C\tp \pm \Gamma\tp,
\eeq
is the optimal Kalman gain, where the minus version will appear shortly,
and the vector of innovations is defined as $\dd\bw\c = \by\dd t - C\bx\c\dd t$ for a given conditioning 
${\rm c} \in \{{\rm F, R, S}\}$, representing filtering, retrofiltering and smoothing, respectively, and satisfies similar 
conditions to \erf{Noise_cond}.

One can also calculate the smoothed estimate of the state 
$\wp\sm(\bcx) \equiv \wp(\bcx|\both{\bf O}) = g(\bcx;\bx\sm,V\sm)$ by conditioning the estimate on the 
past-future measurement record $\both{\bf O}$. The optimal smoothed mean 
$\bx\sm:=\mathbb{E}_{|\both{\bf O}}\{\bcx\}$ and covariance matrix 
$V\sm = \mathbb{E}_{\bcx|\both{\bf O}}\{\bcx\bcx\tp\} - \bx\sm\bx\sm\tp$, can be 
calculated in two ways. The first method is a maximum likelihood 
argument and results in the RTS smoothing equations \cite{Rauch63, RTS65,BLP79},
\begin{align}
\dd\bx\sm &= A\bx\sm\dd t + \tilde{D}V\inv\fil(\cxs - \cxf)\dd t + \Gamma\tp \dd\bw\sm\,,\label{xs-RTS}\\
\ddt{V\sm} &= (\tilde{A} + \tilde{D}V\fil\inv)V\sm + V\sm (\tilde{A} + \tilde{D}V\fil\inv)\tp - \tilde{D}\,,
\label{Vs-RTS}
\end{align}
with the final conditions $\bx\sm(T) = \bx\fil(T)$ and $V\sm(T) = V\fil(T)$. Here, 
$\tilde{A} = A - \Gamma\tp C$, $\tilde{D} = D - \Gamma\tp\Gamma$.

The second method arises from a Bayesian argument, where one first introduces a retrofiltered state that 
runs backwards in time from the final estimation time. This retrofiltered 
state will also be Gaussian with mean $\bx\rfil := \mathbb{E}_{|\fut{\bf O}}\{\bcx\}$ and covariance 
$V\rfil = \mathbb{E}_{\bcx|\fut{\bf O}}\{\bcx\bcx\tp\} - \bx\rfil\bx\rfil\tp$, given by 
\begin{align}
&-\dd\bx\rfil = -A\bx\rfil\dd t + \K^-[V\rfil]\dd\bw\rfil\,,\label{cxr}\\
&-\ddt{V\rfil} = -AV\rfil - V\rfil A\tp + D - \K^-[V\rfil]\K^-[V\rfil]\tp,\label{Vr}
\vspace{-5mm}
\end{align}
with $V\inv\rfil(T) = 0$ describing an uninformative state. 
Combining the filtered and retrofiltered state gives the smoothed state, with the 
mean and covariance described by the MFP smoothing equations \cite{Mayne66,Fraser67,FraPot69}
\begin{align}
&\bx\sm = V\sm[V\inv\fil\bx\fil + V\inv\rfil \bx\rfil]\,,\label{xs-MFP}\\
&V\sm = (V\inv\fil + V\inv\rfil)\inv\,.\label{Vs-MFP}
\end{align}
It has since been shown \cite{FraPot69} that the RTS and MFP forms of the smoothed mean and covariance 
are identical and is easily verified by differentiating \erfs{xs-MFP}{Vs-MFP} with respect to time. 

\subsection{Quantum}
To begin, let us consider an open quantum system whose density matrix $\rho$, assuming a 
Markovian system, evolves according to the Lindblad master equation \cite{WisMil10}
\beq
\hbar\frac{\dd \rho}{\dd t} = -[H, \rho]+ \cal{D}[\hat{\bf c}]\rho\,,
\eeq
with initial condition $\rho(t_0) = \rho_0$. Here $H$ is the systems Hamiltonian describing unitary evolution, 
$\hat{\bf c} = (\hat{c}_1, \hat{c}_2, ..., \hat{c}_K)$ is the vector of Lindblad operators describing the interaction 
between the system and environment, $[A,B] = AB - BA$ is the commutator and the superoperator 
\beq
{\cal D}[\hat{\bf c}]\rho = \sum_{k = 1}^K \hat{c}_k \rho \hat{c}_k\dg - \{\hat{c}_k\dg \hat{c}_k, \rho\}/2\,,
\eeq
with $\{A,B\} = AB + BA$ being the anticommutator. 

For this quantum system to be analogous to a classical LG system, we require that the systems observables 
have an unbounded spectrum. Thus, we will assume that the quantum system that can be described by $N$ 
bosonic modes with each mode described by a position $\hat{q}_k$ and momentum $\hat{p}_k$, which satisfy 
the commutation relation $[\hat q_k,\hat p_\ell] = i\hbar\delta_{k,\ell}$. We can then construct a $2N$ vector 
$\hbx = (\hat q_1,\hat p_1,...,\hat q_N,\hat p_N)\tp$ describing all the modes of the bosonic system. For the 
system to be classified as a LGQ system \cite{DohJac99,WisDoh05,WisMil10}, the Wigner function, a 
quasiprobability distribution defined as 
\beq\label{Wigner}
W(\bcx) = (2\pi)^{-2N}\int \dd^{2N} {\bf b}\, \Tr[\rho {\rm e}^{i\bf{b}\tp(\hbx - \bcx)}]\,,
\eeq
must initially be Gaussian and remain Gaussian, where the latter can be guaranteed when the Hamiltonian is 
quadratic and the vector of Lindblad operators is linear in $\hbx$, i.e., $H = \hbx\tp G \hbx/2$ and 
$\hat{\bf c} = B\hbx$, respectively. Note, since we are using the Wigner function to characterize the 
quantum state, the Hamiltonian must be symmetrically ordered in $\hat{q}_k$ and $\hat{p}_k$, forcing $G$ to 
be a symmetric matrix. As the Wigner function is Gaussian, i.e., $W(\bcx) = g(\bcx; \ex{\hbx}, V)$, all that is 
required to know the state is the mean $\ex{\hbx}$ and covariance matrix, defined symmetrically, 
$V_{k\ell} = \ex{\hat{x}_k\hat{x}_\ell + \hat{x}_\ell\hat{x}_k}/2 - \ex{\hat{x}_k}\ex{\hat{x}_\ell}$, where the 
quantum expectation is $\ex{\bullet} = \Tr[\bullet \rho]$. Using the Lindbald master equation, we can compute 
the dynamic equations for the unconditioned mean and covariance,
\begin{align}
&\dd\ex{\hbx} = A\ex{\hbx}\dd t\,,\\
&\ddt{V} = AV + VA\tp + D\,,
\end{align}
with initial conditions $\ex{\hbx}(t_0) = \ex{\hbx}_0$ and $V(t_0) = V_0$.
Here
\beq
A = \Sigma(G + {\rm Im}[B\dg B]) \qquad {\rm and}\qquad 
D = \hbar\Sigma{\rm Re}[B\dg B]\Sigma\tp\,,
\eeq 
with $\Sigma_{k\ell} = -i[\hat{x}_k,\hat{x}_\ell]/\hbar$ being a symplectic matrix.
Additionally, since the position and momentum operators do not commute, it is necessary that the covariance 
matrix obeys the \SHUR
\beq\label{SHUR}
V + \frac{i\hbar}{2}\Sigma \geq 0\,.
\eeq   
Without any measurement information, this would be the most accurate estimate of the state possible.

If one wishes to obtain a better estimate on the quantum state, it is necessary to gather more information about 
the system by measuring the environment. In the event of a continuous monitoring, we can condition the 
evolution of the state on the past measurement results and obtain the, so 
called, quantum filtering equation \cite{WisMil10, ChiWis11}
\beq\label{Filt_ME}
\hbar\dd \rho\fil = -i[H,\rho\fil]\dd t + {\cal D}[\hat{\bf c}]\rho\fil\dd t + 
\sqrt{\hbar} \dd{\bf w}\fil\tp{\cal H}[M\dg\hat{\bf c}]\rho\fil\,,
\eeq
with $\rho\fil(t_0) = \rho_0$. 
For reasons that will become apparent, I have restricted the measurements to diffusive type 
measurements, like homodyne or heterodyne schemes, with the matrix $M$ (assumed time independent for 
simplicity) characterizes the particular unravelling. Here the filtered innovation is a vector of independent 
Wiener increments defined by $\dd\bw\fil = \by\dd t - \ex{M\dg \hat{\bf c} + \hat{\bf c}\dg M\tp}\fil\dd t$, satisfying 
conditions similar to \erf{Noise_cond}, with the 
conditioned expectation value defined as $\ex{\bullet}\c = \Tr[\bullet \rho\c]$ with ${\rm c} \in\{{\rm F, T, S}\}$, 
representing the filtered, true and smoothed states, respectively, and $\by\dd t$ being the 
measurement current, and the superoperator 
\beq
\begin{split}
{\cal H}[M\dg\hat{\bf c}]\rho = \sum_{k = 1}^{2N} (M\dg)_{k,\ell}&\hat{c}_\ell + \hat{c}\dg_\ell M_{\ell,k} 
\\ &- \Tr[(M\dg)_{k,\ell}\hat{c}_\ell + \hat{c}\dg_\ell M_{\ell,k}]\,,
\end{split}
\eeq
where the Einstein summation convention is being used over repeated indices. To ensure that the resulting filtered 
state is a valid quantum state, it is necessary and sufficient \cite{ChiWis11} that the matrix $M$ satisfies 
$MM\dg = {\rm diag}(\eta_1, \eta_2, ..., \eta_{2N})$, where $0 \leq \eta_k \leq 1\, \forall\, k$ and can be 
interpreted as the fraction of the channel $\hat{c}_k$ that has been observed. Note, it must be the case that 
at least one $\eta_k > 0$, otherwise no measurement has been performed and both filtering and 
smoothing are redundant.

When considering an LGQ system, in order to keep the filtered Wigner function Gaussian throughout the entire 
evolution we require the measurement current to be linear in $\hbx$ (which is the case for diffusive type 
measurements), i.e., 
\beq
\by\dd t = C\hbx\dd t + \dd\bv\m\,,
\eeq
where the measurement matrix $C = 2\sqrt{\hbar\inv} T\tp \tilde{B}$,
$T\tp = [{\rm Re}(M\tp), {\rm Im}(M\tp)]$, $\tilde{B}\tp = [{\rm Re}(B\tp), {\rm Im}(B\tp)]$, and the measurement 
noise, for simplicity, is assumed to be white, i.e., satisfies the properties in \erf{Noise_cond}. As was the case 
for the unconditioned state, since the filtered state has been restricted to have a Gaussian Wigner function,
$W\fil(\bcx) = g(\bcx;\xfil,V\fil)$, we only need information about the mean and covariance matrix to specify the 
state. Using \erf{Filt_ME}, on can obtain the equations for the filtered mean and covariance matrix 
\cite{DohJac99,WisDoh05,WisMil10}. 
\begin{align}
&\dd\xfil = A\xfil\dd t + \K^+[V\fil]\dd\bw\fil\,,\label{xfil}\\
&\ddt{V\fil} = AV\fil + V\fil A\tp + D - \K^+[V\fil]\K^+[V\fil]\tp\,, \label{Vfil}
\end{align}
with initial conditions $\xfil(t_0) = \ex{\hbx}_0$ and $V\fil(t_0) = V_0$. Here $\K^+[V]$ is defined in \erf{gain}, 
with $\Gamma = -\sqrt{\hbar} T\tp S\tilde{B}\Sigma\tp$ and $S = [\begin{smallmatrix}
0 & I_K\\
-I_K & 0
\end{smallmatrix}]$.
Note, since the filtered state $\rho\fil$ is a 
valid quantum state, it is the case that $V\fil$ satisfies the \SHUR.

If one wishes to obtain an even more accurate estimate of the state, it 
is possible that we can condition the estimate of the state on not only the past measurement information but 
also the future measurement record. This leads us to the quantum state smoothing theory of Guevara and 
Wiseman \cite{GueWis15}. 

\section{Deriving the Quantum Rauch-Tung-Striebel Smoothed State} \label{Sec-Derivation}
In order to calculate a smoothed quantum state, one needs to first introduce the true state $\rho\god$. As 
mentioned in the introduction, the true state is an estimate conditioned on two independent measurement 
records, the past record observed by Alice and the past record observed by Bob. This situation is merely an  
extension of the filtering scenario to multiple independent measurement records, and thus it is simple to 
see that the resulting stochastic master equation will be 
\beq
\begin{split}
\hbar\dd\rho\god = -i[H,\rho\god]&\dd t + {\cal D}[\hat{\bf c}]\rho\god\dd t \\& + 
\sum_{{\rm r} \in\{{\rm o}, {\rm u}\}}\sqrt{\hbar} \dd{\bf w}_{\rm r}\tp{\cal H}[M_{\rm r}\dg\hat{\bf c}]\rho\god\,,
\end{split}
\eeq
where, assuming Alice and Bob have the same initial information about the quantum state, the initial condition 
is $\rho\god(t_0) = \rho_0$. Note, this may not always be the case and is trivial to adapt to the general case, 
however this assumption seems like a scenario that would occur frequently and is worth considering. As before, 
we are restricting to the case where both Alice and Bob measure 
there respective fractions of the measurement channels using a diffusive-type measurements. We have 
introduced the (r)ecord subscript to distinguish between the record (o)bserved by Alice and the record 
(u)nobserved by Alice (Bob's record), where the matrices $M\ob$ and $M\un$ describe how Alice and Bob have 
unravelled the the system, respectively. The observed and unobserved innovations are defined as 
$\dd\bw_{\rm r} = \by_{\rm r}\dd t - \ex{M\dg_{\rm r} \hat{\bf c} + \hat{\bf c}\dg M_{\rm r}\tp}\god\dd t$ where we 
have assumed both records are uncorrelated, i.e., $\dd\bw\ob\dd\bw\un\tp = 0$ and $\by_{\rm r}\dd t$ is 
corresponding the measurement current for Alice and Bob. 

As was the case for the filtered state, to ensure that 
the true state is a valid quantum case it is necessary and sufficient that both $M\ob$ and $M\un$ must satisfy 
$M_{\rm r} M_{\rm r}\dg = {\rm diag}(\eta_{{\rm r},1}, \eta_{{\rm r},2}, ..., \eta_{{\rm r},M})$ with 
$0 \leq \eta_{\rm r, k} \leq 1$ and $\eta_{{\rm o},k} + \eta_{{\rm u},k} \leq 1\,\forall k$. Note, although it is often 
convenient to assume that $M\ob M\ob\dg + M\un M\un\dg = I_M$ \cite{LCW19,LCW-QS20,LCW-PRA21}, 
that is, together Alice's and Bob's 
measurements constitute a perfect measurement of the system and the resulting true state will be pure, it not a 
necessary assumption and what follows will hold generally. As was the case with filtering, it must be that 
at least one $\eta_{{\rm o}, k} > 0$ and $\eta_{{\rm u},k'} > 0$.

For an LGQ system, to ensure that the true state remains Gaussian, as was the case for filtering, we require that 
both Alice's and Bob's measurements are linear,
\beq
\by\rr\dd t = C\rr\xgod\dd t + \dd\bw\rr\,,
\eeq
where, for simplicity, I have defined the currents in terms of the true mean $\xgod$ and the innovation, where, 
in this form, it should be clear that observed innovation 
$\dd\bw\ob = \by\ob\dd t - C\ob\xgod\dd t$ is different from Alice's filtered innovation 
$\dd\bw\fil = \by\ob\dd t - C\ob\xfil\dd t$. 
Given this restriction, the Wigner function for the true state will be 
Gaussian $W\god(\bcx) = g(\bcx,\xgod,V\god)$ with the true mean $\xgod$ and covariance $V\god$
\cite{LCW19,LCW-PRA21}
\begin{align}
&\dd\xgod = A\xgod\dd t + \sum_{{\rm r} \in\{{\rm o},{\rm u}\}} \K_{\rm r}^+[V\god]\dd\bw_{\rm r},\label{xt}\\
&\ddt{V\god} = AV\god + V\god A\tp + D  - \sum_{{\rm r} \in\{{\rm o},{\rm u}\}}\K_{\rm r}^+[V\god]
\K_{\rm r}^+[V\god]\tp.\label{Vt}
\end{align}
with initial conditions $\xgod(t_0) = \ex{\hbx}_0$ and $V\god(t_0) = V_0$.
Here, $\K\rr^\pm[V] = V C\rr\tp \pm \Gamma\rr\tp$, where $C_{\rm r}$ and $\Gamma_{\rm r}$ are defined in in 
the same way as before, with the appropriate $M_{\rm r}$ used in both cases.

Now that the true state has been calculated, we can begin to derive the smoothed quantum state for LGQ 
systems. To begin, as we are interested in the Wigner function of the smoothed state $W\sm(\bcx)$ in the LGQ 
setting, we can apply \erf{Wigner} to both sides of \erf{qsm}, where, by the linearity of the trace, we obtain 
\beq\label{Sm_Wig}
W\sm(\bcx) = \mathbb{E}_{\past{\bf U}|\both{\bf O}}\{W\god(\bcx)\}\,.
\eeq
As the Wigner function of the true state is restricted to a Gaussian, we know that it only depends on the mean 
$\xgod$ and covariance matrix $V\god$, the latter of which is deterministic with the former depending explicitly 
on both $\past{\bf O}$ and $\past{\bf U}$. Hence, averaging over $\past{\bf U}$ with a fixed $\both{\bf O}$ will 
be equivalent to averaging over the true mean $\xgod$ for a fixed observed record. To make the notation 
simpler for the derivation to come, we take $\halo\bx = \xgod$, where the ellipse accent will be refereed to as a 
`halo' and will denote intermediary variables between the true state and the filtered/smoothed state. Making 
these changes to \erf{Sm_Wig}, we get
\beq\label{Sm_Wig_2}
W\sm(\bcx) = \mathbb{E}_{\halo{\bf x}|\both{\bf O}}\{W\god(\bcx)\} = \int \dd\mu(\halo\bx) \wp(\halo\bx|\both{\bf O}) g(\bcx; 
\halo\bx, V\god)\,,
\eeq
where the integral measure $\dd\mu(\halo\bx) = \prod_{k = 1}^{2N}\dd\halo{x}_k$. We also know that 
$\wp(\halo\bx|\past{\bf O})$ will be a Gaussian distribution. To see why this is the case, we rewrite \erf{xt} as
\beq\label{halo_uncond}
\dd\halo{\bx} = A\halo{\bx}\dd t + \bar{E}\dd\bar\bv\p\,,
\eeq
where $\bar{E}\dd\bar{\bv}\p = \sum_{{\rm r} \in\{{\rm o},{\rm u}\}} \K_{\rm r}^+[V\god]\dd\bw_{\rm r}$, with the 
observed measurement current
\beq
\by\ob\dd t = C\ob\halo{\bx}\dd t + \dd\bw\ob\,,
\eeq
where it is clear that $\halo\bx$ is described by a linear Langevin equation, as in \erf{LLE}, and by conditioning 
on a linear measurement current the resulting PDF will be Gaussian. 

As $\wp(\halo\bx|\both{\bf O})$ is a classical object, we can simply apply classical smoothing 
theory in order to compute this PDF, and since this PDF is Gaussian, 
$\wp(\halo\bx|\both{\bf O}) = g(\halo{\bx}; \halo\bx\sm, \halo{V}\sm)$, we only need to determine the mean 
$\halo\bx\sm$ and covariance $\halo V\sm$. At this point there are two paths we can take, we can use either 
the MFP or RTS smoothing technique on \erf{halo_uncond} to obtain equations for the haloed smoothed mean 
and covariance. As stated earlier, I will choose the latter. Applying \erfs{xs-RTS}{Vs-RTS} to the 
\erf{halo_uncond} results in 
\begin{align}
&\dd\halo\bx\sm = A\halo\bx\sm\dd t + \bar{D}\hV\fil\inv(\halo\bx\sm-\halo\bx\fil)\dd t  
+ \bar{\Gamma}\tp\dd\bw\sm\,,\label{dhxs}\\
&\ddt{\hV\sm} = (\bar{A} +\bar{D}\hV\fil\inv)\hV\sm + \hV\sm (\bar{A} +\bar{D}\hV\fil\inv)\tp - \bar{D}\,, \label{dhVs}
\end{align}
with final conditions $\halo\bx\sm(T) = \halo\bx\fil(T)$ and $\halo V\sm(T) = \halo V\fil(T)$.
Here $\bar{A} = A - \bar{\Gamma}\tp C\ob$, $\bar{D} = \bar{E}\bar{E}\tp - \bar\Gamma\tp\bar\Gamma$, 
$\bar\Gamma\tp \dd t = \bar{E}\dd\bar\bv\p\dd\bw\ob\tp = \K\ob^+[V\god]\dd t$ and 
$\dd\bw\sm = \by\ob\dd t - C\ob\halo\bx\sm\dd t$. Additionally, I have introduced both 
the haloed filtered mean $\halo\bx\fil = \mathbb{E}_{|\past{\bf O}}\{\halo{\bx}\}$ and covariance 
$\halo V\fil = \mathbb{E}_{\halo{\bx}|\past{\bf O}}\{\halo{\bx} \halo{\bx}\tp\} - \halo{\bx}\fil\halo{\bx}\fil\tp$ which are 
obtained from the filtered PDF $\wp(\halo\bx|\past{\bf O})$. Note, in \erfs{dhxs}{dhVs} I have assumed that 
$\halo{V}\fil$ is invertible, which may not always be the case. In the event that $\hV\fil$ is not invertible the 
smoothed quantum state can still be computed with slight modifications, see Appendix \ref{App1} for details. 

While the haloed filtered mean and covariance do satisfy there own 
differential equations, it has been shown \cite{LCW19} that $\halo\bx\fil = \xfil$ and $\halo V\fil = V\fil - V\god$, 
thus the specific equations are not important for computing the haloed smoothed mean and covariance 
matrices. However, for simplicity, the haloed covariance matrix will be used more often for simplicity. 
It should also be emphasized that $\halo{V}\fil$ is the covariance 
of a classical state and thus is not required to satisfy the \SHUR, unlike $V\fil$ and $V\god$.

Finally, we have all the necessary information to compute the smoothed quantum state for LGQ systems. 
Returning to \erf{Sm_Wig_2}, we can substitute in the Gaussian PDF, 
$\wp(\halo{\bx}|\both{\bf O}) = g(\halo\bx; \halo\bx\sm, \halo V\sm)$, obtaining
\beq\label{final_sm_int}
g(\bcx; \xsm, V\sm) = \int \dd\mu(\halo\bx) g(\halo\bx; \halo\bx\sm, \halo V\sm) g(\bcx; \halo\bx, V\god)\,,
\eeq
where, using the fact that convolving two Gaussian functions will result in another Gaussian, I have replaced 
the Wigner function of the smoothed state with its Gaussian with mean $\xsm$ and covariance matrix $V\sm$. 
Using the properties of such a convolution, we find that $\xsm = \halo\bx\sm$ and 
$V\sm = \halo V\sm + V\god$. Thus, using $\ddt{V\sm} = \ddt{\halo V\sm} + \ddt{V\god}$, we obtain the RTS 
form of the quantum state smoothing equations, 
\begin{align}
&\dd\xsm = A\xsm\dd t + \bar{D}\hV\fil\inv(\xsm-\xfil)\dd t  + \K\ob^+[V\god]\dd\bw\sm\,,\label{dxs}\\
&\ddt{V\sm} = (\bar{A} +\bar{D}\hV\fil\inv)V\sm + V\sm(\bar{A} +\bar{D}\hV\fil\inv)\tp + Q\,,\label{dVs}
\end{align}
with the final condition $V\sm(T) = \hV\fil(T) + V\god(T) = V\fil(T)$ and 
\beq
Q = D - \Gamma\ob\tp\Gamma\ob + V\god C\ob\tp C\ob V\god - \bar{D}\hV\fil\inv V\god - V\god\hV\fil\inv\bar{D} - 2\bar{D}\,.
\eeq

\section{Differentiability of the Quantum Smoothed Mean}\label{Sec-Disc}
In addition to being simpler to compute, as one does not need to calculate the haloed retrofiltered estimates 
\cite{LCW19, LCW-PRA21}, the RTS forms of the smoothed mean and covariance, \erfs{dxs}{dVs}, are dynamical 
equations which can provide insight into how the smoothed state evolves. This is hidden in the MFP forms. 
In particular, we can see the 
non-differentiability of the smoothed mean. This may not be terribly surprising as even in the classical case the 
smoothed mean was non-differentiable. However, there is a slight, perhaps predictable difference 
between the innovation terms in the classical and quantum cases. Specifically, we see that, in the quantum 
case, there is a dependence on the true covariance $V\god$, which is reasonable as this is the minimum 
uncertainty in the mean for the quantum system given Alice's and Bob's measurements. Thus it is reasonable to 
expect that Alice's measurement would change her estimate of the mean by at least an amount proportional to 
$V\god$. In contrast, the innovation in the classical case only depends on the cross-correlation matrix 
$\Gamma$ since the true covariance is zero, corresponding to a delta function PDF, where the \SHUR~prevents 
$V\god\to 0$ in quantum systems. Furthermore, because of the uncertainty relation, one 
might expect that the quantum smoothed mean will always be non-differentiable. This is not the case.

In the classical case, all that is required for the smoothed mean $\bx\sm$ to be differentiable is that 
the cross-correlation matrix $\Gamma = 0$, meaning that Alice's measurement is uncorrelated with the noise 
affecting the system. However, this condition is not sufficient in general for the quantum case. To find the condition 
for differentiability in the quantum case, we will assume that the mean is differentiable, that is, 
$\dd\xsm\propto\dd t$, over the interval $[\tau_1, \tau_2]$. It is easy to see by looking at \erf{dxs}, the only term 
that is preventing the mean from 
being differentiable is the innovation term $\K\ob^+[V\god]\dd\bw\sm$. Thus, in order for the mean to be 
differentiable, this innovation term must either be proportional to $\dd t$ or vanish over the interval. It is impossible 
for the former to be true as for the innovation term to be proportional to $\dd t$, since 
$\dd\bw\sm\tp\dd\bw\sm = I_{2N}\dd t$, it must be the case that $\K\ob^+[V\god] = R\dd\bw\sm\tp$, where $R$ is 
an arbitrary matrix, 
which cannot be the case as $V\god$, $C\ob$ and $\Gamma\ob$ are all independent of the observed 
measurement at the estimation time. The latter, on the other hand, is not impossible and occurs when 
$\K\ob^+[V\god] = 0$. Since we are considering a fixed measurement scheme, i.e., $C\ob$ and $\Gamma\ob$ are 
time-independent, it is impossible for this condition to be satisfied at all times. Thus, to have a differentiable 
smoothed mean over a non-infinitesimal time interval, I will only consider time intervals in the steady state regime, 
i.e., $\tau_1 \geq \tau^{\rm ss}$, where $\tau^{\rm ss}$ is the time taken for the true covariance to reach steady 
state.

Under this condition, the steady-state of the true covariance $V^{\rm ss}\god$ satisfies
\beq
0 = AV^{\rm ss}\god + V^{\rm ss}\god A\tp + D - \K\un^+[V^{\rm ss}\god]\K\un^+[V^{\rm ss}\god]\tp\,.
\eeq
At this point we can see that the steady-state covariance of the true state is identical to the steady-state of a single 
record filtered state, like \erf{Vfil}, however, rather than using Alice's past record, it is Bob's past record that is 
used. For comparison, Bob's filtered state is 
$W_{\past{\bf U}}(\bcx) = g(\bcx; \ex{\hbx}_{\past{\bf U}}, V_{\past{\bf U}})$, with
\begin{align}
&\dd\ex{\hbx}_{\past{\bf U}} = A\ex{\hbx}_{\past{\bf U}}\dd t + \K\un^+[V_{\past{\bf U}}]\dd\bw_{\past{\bf U}}\,,\\
&\ddt{V_{\past{\bf U}}} = AV_{\past{\bf U}} + V_{\past{\bf U}}A\tp + D - \K\un^+[V_{\past{\bf U}}]
\K\un^+[V_{\past{\bf U}}]\,,\label{Vu}
\end{align}
where $\dd\bw_{\past{\bf U}} = \by\un\dd t - C\un \ex{\hbx}_{\past{\bf U}}\dd t$.
Thus, we have that if the mean of the smoothed quantum state is differentiable in the steady-state regime then 
$V\god^{\rm ss} = V_{\past{\bf U}}^{\rm ss}$.

Importantly, the converse of this is also true, that is, if $V\god^{\rm ss} = V_{\past{\bf U}}^{\rm ss}$ then the mean 
of the smoothed quantum state is differentiable. This is simple to see since the steady-state of the true covariance in general satisfies
\beq\label{Vt_ss}
\begin{split}
0 = A&V^{\rm ss}\god + V^{\rm ss}\god A\tp + D \\&- \K\ob^+[V^{\rm ss}\god]\K\ob^+[V^{\rm ss}\god]\tp- \K\un^+[V^{\rm ss}\god]\K\un^+[V^{\rm ss}\god]\tp\,,
\end{split}
\eeq
and the steady state of Bob's filtered covariance satisfies
\beq
0 = AV^{\rm ss}_{\past{\bf U}} + V^{\rm ss}_{\past{\bf U}} A\tp + D - \K\un^+[V^{\rm ss}_{\past{\bf U}}]\K\un^+[V^{\rm ss}_{\past{\bf U}}]\tp\,.
\eeq
Since $V\god^{\rm ss} = V_{\past{\bf U}}^{\rm ss}$, \erf{Vt_ss} reduces to 
$\K\ob^+[V^{\rm ss}\god]\K\ob^+[V^{\rm ss}\god]\tp = 0$ giving $\K\ob^+[V^{\rm ss}\god] = 0$. As we have already 
shown under this condition $\dd\xsm \propto \dd t$ and is differentiable. As a result, 
we have the necessary and sufficient condition that $V\god^{\rm ss} = V_{\past{\bf U}}^{\rm ss}$ for a 
differentiable mean of the smoothed quantum state in the steady state regime.

\begin{figure*}[t!]
\begin{minipage}{0.5\textwidth}
\includegraphics[scale = 0.3]{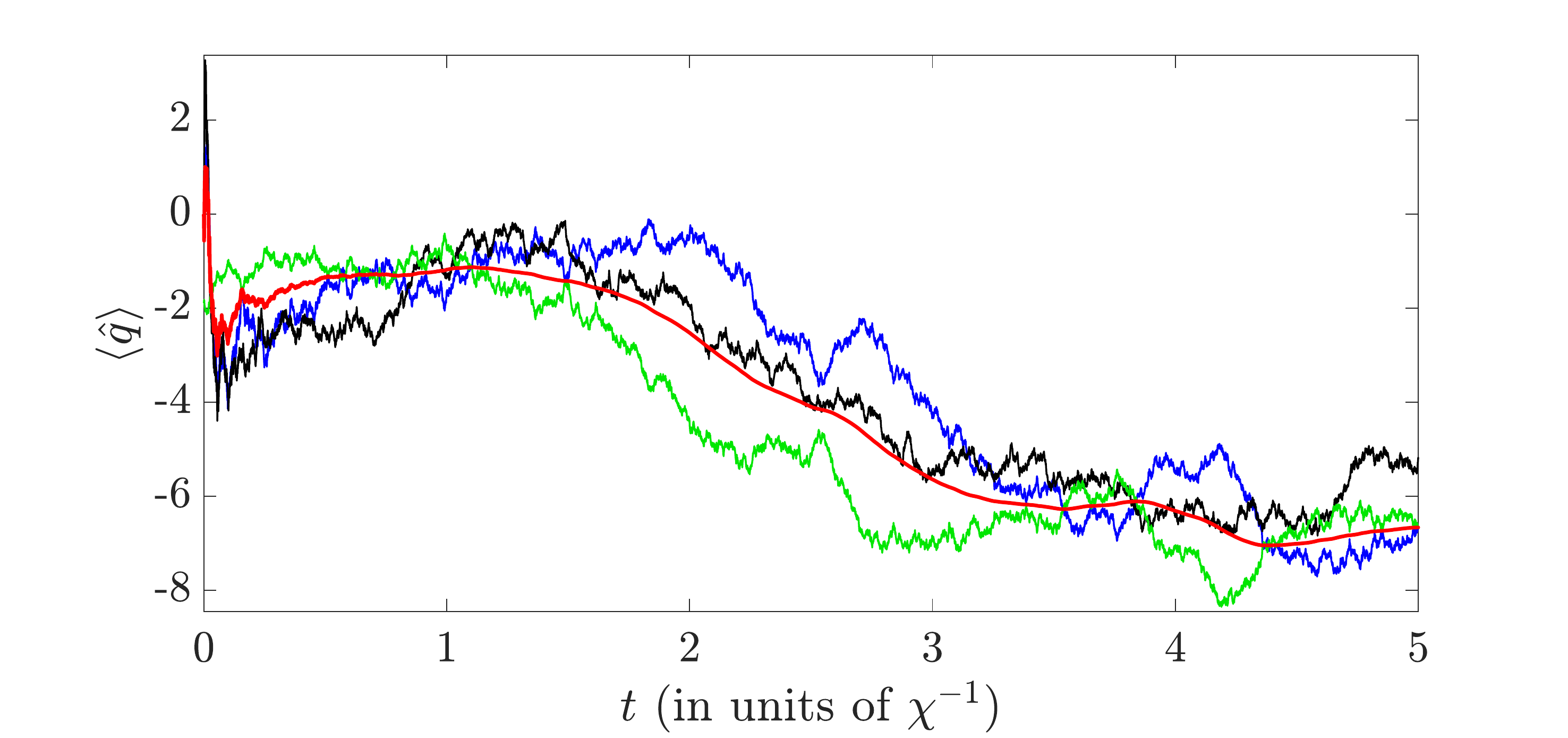}\\
\includegraphics[scale = 0.3]{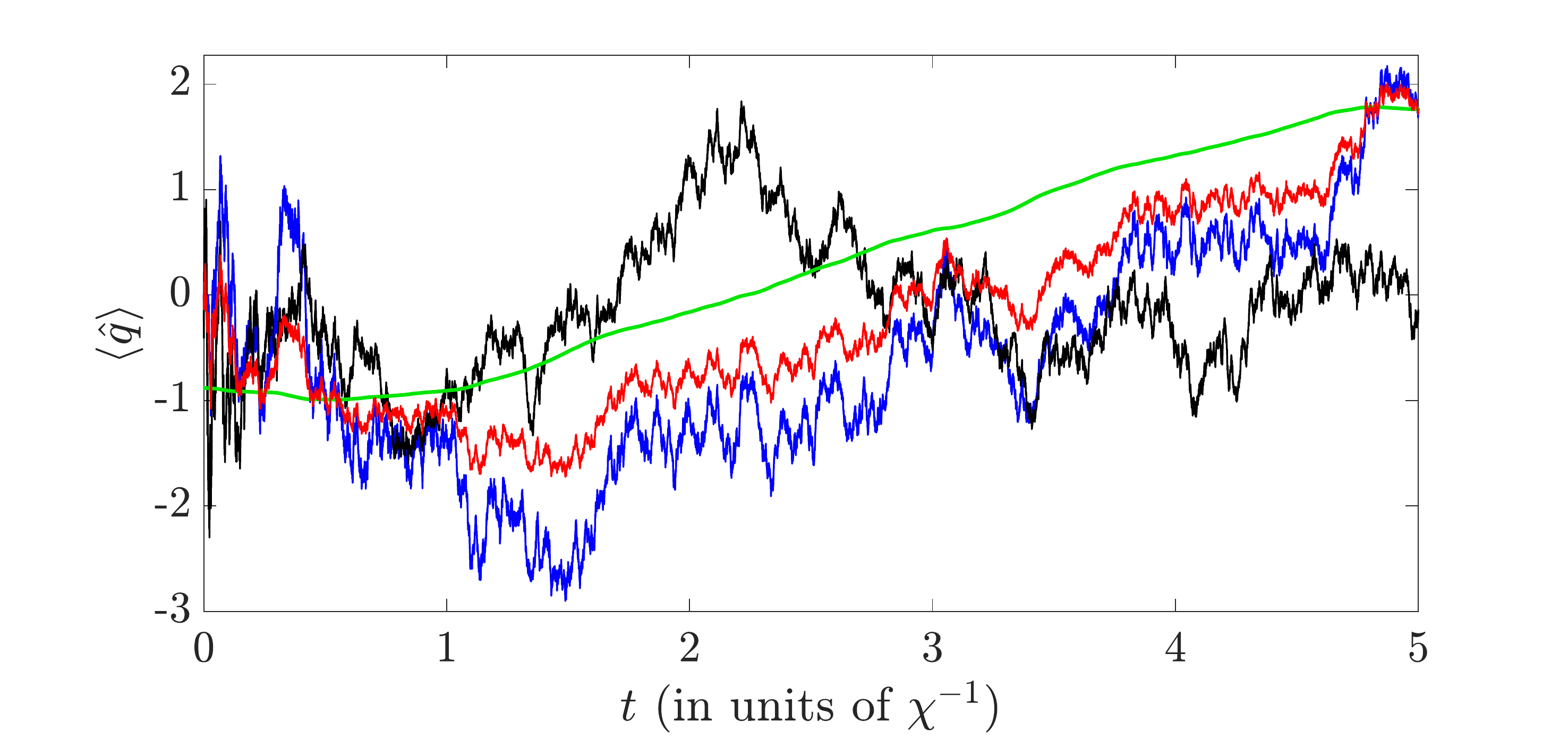}
\end{minipage}%
\begin{minipage}{0.5\textwidth}
\includegraphics[scale = 0.3]{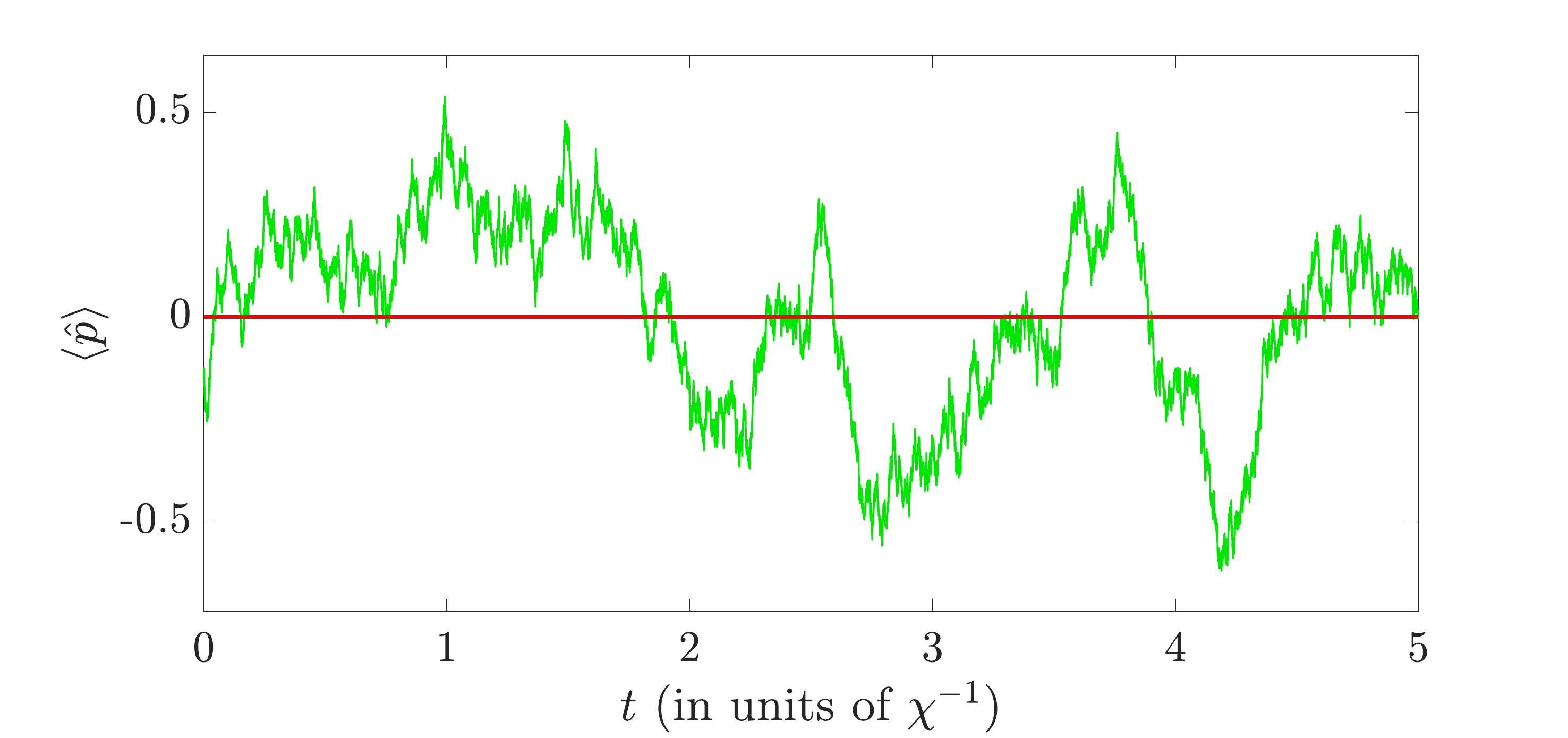}\\
\includegraphics[scale = 0.3]{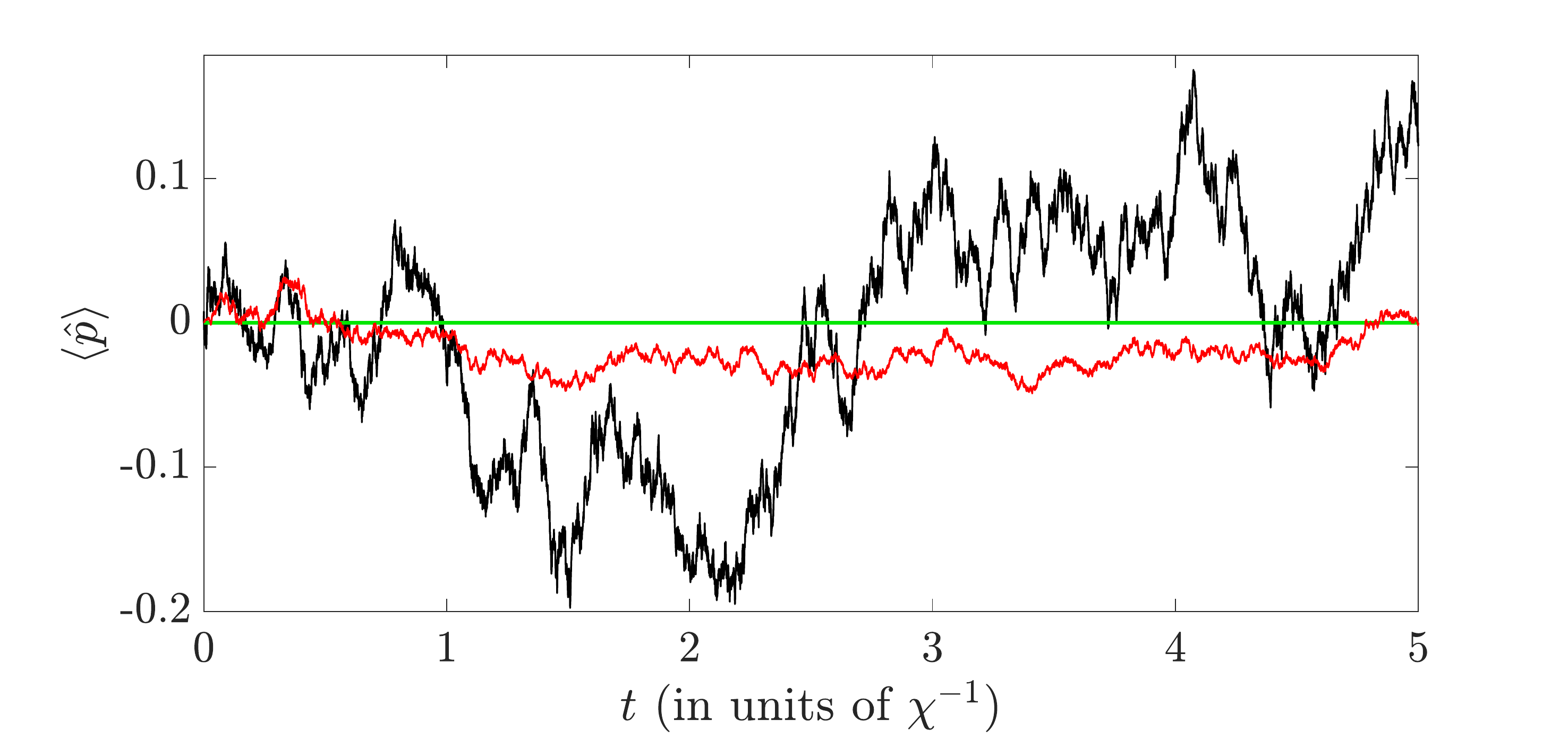}
\end{minipage}
\caption{A single realization of the $q$- and $p$-quadratures (left and right, respectively) for the quantum 
system \erf{LME}. The top two graphs are when Alice is measuring the $\gamma$-channel with homodyne 
phase $\theta_{\rm \gamma, o} = \pi/8$ and Bob is measuring the $\kappa$-channel with $\theta_{\rm \kappa, u} = 0$, where $g = 1$. The quantum smoothed 
mean (red line) initially evolves stochastically. However, at $t \approx 0.8$, a sufficient time has passed for the 
system to reach steady state and the quantum smoothed mean begins to evolve continuously since 
$V\god^{\rm ss} = V^{\rm ss}_{\protect\past{\bf U}}$. Whereas the other three estimates, the true mean (black line), the filtered mean 
(blue line) and the classically smoothed mean (green line), na\"{i}vely applied to the quantum system, all evolve 
stochastically. For this case, $\ex{\hat{p}}\god(t) = \ex{\hat{p}}\fil(t) =0$. 
In the bottom two graphs the reverse scenario is considered for Alice and Bob. that is, Alice measures the 
$\kappa$-channel with $\theta_{\rm \kappa, o} = 0$ and Bob measures the $\gamma$-channel with 
homodyne phase $\theta_{\rm \gamma, u} = \pi/8$, in this 
case with $g = 0.1$. While Alice's measurement has a zero cross-correlation matrix, i.e., $\Gamma\ob = 0$, 
the quantum smoothed mean (red line) evolves stochastically, since 
$V\god^{\rm ss} \ne V^{\rm ss}_{\protect\past{\bf U}}$. I have included 
the true mean (black line) and filtered mean (blue line) for completeness, where $\ex{\hat{p}}\fil(t) = 0$. Also, in 
this case, since Alice's cross-correlation matrix is zero, the classically smoothed mean evolves continuously. In 
both cases, $\hbar = 2$ and the initial conditions are taken to be $\ex{\hbx}_0 = (0,0)\tp$ and 
$V_0 = {\rm diag}(10,(1+g)/2)$, where $V_0$ was chosen to be a finite version of the unconditioned steady 
state.}
\label{Fig-1}
\end{figure*} 

Note, this is true more generally outside the steady-state regime, when 
$V\god(t) = V_{\past{\bf U}}(t)$ over a time interval $[\tau_1, \tau_2]$. To see this, we note that since 
$V\god(t) = V_{\past{\bf U}}(t)$, their derivatives at all times in the interval are also equal. Subtracting \erf{Vt} 
from \erf{Vu} gives $\K\ob^+[V\god]\K\ob^+[V\god]\tp = 0$ and thus $\K\ob^+[V\god] = 0$ for all times in the interval. Hence, in the general case, it is sufficient for a differentiable smoothed mean over the interval 
$[\tau_1, \tau_2]$ if $V\god(t) = V_{\past{\bf U}}(t)$. However, it is not a necessary condition.

This differentiability condition contains useful information about the type of systems and measurements 
that could exhibit a differentiable smoothed mean. Specifically, we see from 
$V\god^{\rm ss} = V_{\past{\bf U}}^{\rm ss}$ that 
Alice's measurement cannot reduce the uncertainty of the true state beyond what Bob's measurement already 
has. This immediately means that the quantum system must have more than one Lindblad operator, otherwise the 
portion of the channel that Alice monitors will always reduce the uncertainty by a fraction more when combine with 
Bob's measurement. 

As an example, I will consider a single mode $(N = 1)$ open quantum system described by the master equation,
\beq\label{LME}
\hbar\ddt{\rho} = -i\chi[(\hat{q}\hat{p} + \hat{p}\hat{q})/2,\rho] + \gamma{\cal D}[\hat{q} + i\hat{p}]\rho + 
\kappa{\cal D}[\hat{q}]\rho\,.
\eeq
Note, while this is a toy system to illustrate the differentiability condition for quantum state smoothing, in 
principle it could be constructed using linear optics. In fact, this system is similar to an optical parametric 
oscillator \cite{WisDoh05, WisMil10}, where the only difference is the additional Lindblad operator $\hat{c}_2 = \hat{q}$, as mentioned before 
is required. In this system, since the variance in the $\hat p$ quadrature is bounded by the squeezing Hamiltonian, 
we can expect that, when $\kappa$ is large enough, the amount of information about the position quadrature 
gained by monitoring $\hat{c}_2$ would reduce the uncertainty in $\hat q$ by enough so that the state is pure. 
Thus, for this system, if Bob where to perfectly monitor $\hat{c}_2$ (the $\kappa$-channel), with Alice perfectly 
monitoring $\hat{c}_1$ (the $\gamma$-channel), then the steady state of the  true state will be pure, as both Alice 
and Bob have performed a perfect monitoring, and the steady state of Bob's filtered state will also be pure with the 
same covariance, satisfying the differentiability condition.

To make this more formal, time will be measured in units of $\chi\inv$ and I will consider the case 
$\gamma = \chi$. For this system, the drift and diffusion matrices are 
$A = {\rm diag}(0,-2)$ and $D = \hbar\cdot{\rm diag}(1, 1+g)$, where $g = \kappa/\chi$. I will assume that 
Alice and Bob both perform homodyne measurements. For homodyne measurements, the matrix 
$M_{\rm r} = {\rm diag}(\sqrt{\eta_{{\rm \gamma},{\rm r}}} \exp[i\theta_{{\rm \gamma},{\rm r}}], 
\sqrt{\eta_{{\rm \kappa},{\rm r}}} \exp[i\theta_{{\rm \kappa},{\rm r}}])$, with the resulting measurement matrix 
\beq
C_{\rm r} = 2\sqrt{\hbar\inv} \left[\begin{array}{cc}
\sqrt{\eta_{\gamma,{\rm r}}} \cos\theta_{\gamma,{\rm r}} & \sqrt{\eta_{\gamma,{\rm r}}}
\sin\theta_{\gamma,{\rm r}}\\
\sqrt{g\eta_{\kappa,{\rm r}}} \cos\theta_{\kappa,{\rm r}} & 0
\end{array}\right]\,,
\eeq
and cross-correlation matrix 
\beq
\Gamma_{\rm r} = -\sqrt{\hbar}\left[\begin{array}{cc}
\sqrt{\eta_{\gamma,{\rm r}}} \cos\theta_{\gamma,{\rm r}} & 
\sqrt{\eta_{\gamma,{\rm r}}}\sin\theta_{\gamma,{\rm r}}\\
0 & \sqrt{g\eta_{\kappa,{\rm r}}} \sin\theta_{\kappa,{\rm r}}
\end{array}\right]\,.
\eeq
Here, $\eta_{n,{\rm r}}$ and $\theta_{n,{\rm r}}$ is the measurement efficiency and homodyne 
phase, respectively, for each channel $n \in \{\gamma,\kappa\}$ and observer ${\rm r} \in \{{\rm o, u}\}$. 

To demonstrate the smooth evolution of the quantum smoothed mean, Bob must measure the position 
quadrature of the $\kappa$-channel perfectly, i.e.  $\eta_{\gamma,{\rm o}} = 1$ and
$\theta_{\gamma,{\rm o}} = \pi/8$, $\eta_{\kappa,{\rm u}} = 1$ and $\theta_{\kappa,{\rm u}} = 0$. Crucially, $g$ 
is chosen to be unity as this is the point where the steady state solution for Bob's filtered covariance is pure and is 
equal the true state. Alice, on the other hand, perfectly monitors the $\gamma$ channel, i.e., 
$\eta_{\kappa,{\rm u}} = 1$, with some homodyne phase $\theta_{\kappa,{\rm u}} = \pi/8$. Note, the particular 
homodyne phase for Alice does not matter in the slightest for the theory and was just chosen for the 
simulations. As the top graphs of 
Fig.~\ref{Fig-1} show, after a sufficient time ($t \approx 0.8$) for the true covariance to reach its steady 
state, the quantum smoothed mean (red line) begins to evolve smoothly, as one would expect from a 
differentiable function. Whereas, the other three estimates, i.e. the true, the filtered and the classically 
smoothed means, all evolve stochastically over the entire time interval. The classically smoothed mean was 
computed using \erfs{xs-RTS}{Vs-RTS}, where the measurement 
matrix $C$ and the cross-correlation matrix $\Gamma$ have been replaced with Alice's measurement matrix 
$C\ob$ and cross-correlation matrix $\Gamma\ob$.

Now, if we reverse the channels that Alice and Bob measure, i.e., $\eta_{\gamma,{\rm u}} = 1$, 
$\theta_{\gamma, {\rm u}} = \pi/8$ and  $\eta_{\kappa,{\rm o}} = 1$, $\theta_{\kappa, {\rm o}} = 0$, we would 
expect the smoothed mean to be non-differentiable throughout the entire evolution, like the filtered and true state. 
This is clearly the case, as seen in the bottom graphs of Fig.~\ref{Fig-1}. Note, for this case $g \ne 1$ so that 
Alice's filtered and smoothed state do not reduce to the true state when the system reaches steady state. 

The fact that the quantum smoothed mean evolves stochastically might bring into question 
whether this technique should be referred to as smoothing. While it is idiosyncratic that a `smoothing' technique 
does not provide a smooth estimate, I reiterate that the classically smoothed mean suffers from the same issue 
when $\Gamma\ob \ne 0$, as seen in Fig.~\ref{Fig-1}. I also remind the reader that the smoothing technique 
refers to using future measurement information as well as the past information to obtain an estimate 
\cite{Weinert01, Hay01, vanTrees13, BroHwa12, Einicke12, Fri12, Sarkka13} and not necessarily obtaining a 
smooth estimate. Thus, I believe there is no issue in referring to this quantum state estimation technique as 
smoothing.

\section{Conclusion}
In this paper I have derived the Rauch-Tung-Striebel forms of the quantum state smoothing equations for LGQ 
systems. These new forms not only make it easier to compute the smoothed quantum state 
but also provide insight into the dynamics of the state. From these equations, I have derived a necessary and 
sufficient condition for the quantum smoothed mean to be differentiable in the steady state limit. These equations 
could prove very useful in the future in identifying more properties of the smoothed state. A good example in the 
existing literature could be to provide a explanation or even an analytic solution for the optimal measurement 
strategies for Alice and Bob presented in Ref.~\cite{LCW-PRA21}.
Additionally, there may even be uses for this smoothing 
technique outside of quantum mechanics. This result would hold for any classical linear Gaussian system with 
a minimum bound on the covariance, i.e. a system with an uncertainty relation. Lastly, it would be interesting to 
see if the sufficient condition $V\god(t) = V_{\past{\bf U}}(t)$ for a differentiable quantum smoothed mean is 
sufficient outside the LGQ regime.

\begin{acknowledgments}
I would like to thank Howard M. Wiseman for his invaluable comments and the example that illustrates 
the differentiability of the quantum smoothed mean. 
I acknowledge the traditional owners of the land on which this work was undertaken at Griffith 
University, the Yuggera people. This research is supported by an Australian Government 
Research Training Program (RTP) Scholarship.
\end{acknowledgments}

\appendix
\section{Non-invertible haloed filtered covariance}\label{App1}
In order to show how to treat the smoothed quantum state when $\hV\fil\inv$ does not exist,
I will go back to the more general definitions and consider the definition of the filtered state in terms of the true 
state, that is, 
$\rho\fil = \mathbb{E}_{\past{\bf U}|\past{\bf O}}\{\rho\god\}$ \cite{GueWis15}. Following similar steps to 
the derivation of the smoothed quantum state in Sec.~\ref{Sec-Derivation}, arriving at
\beq
W\fil(\bcx) = \int\dd\mu(\halo{\bx}) \wp(\halo{\bx}|\past{\bf O})W\god(\bcx)\,.
\eeq
In particular, I will look at the probability distribution 
$\wp(\halo{\bx}|\past{\bf O}) = g(\halo{\bx};\halo{\bx}\fil,\halo{V}\fil)$, where the Gaussianity follows from a similar 
argument to $\wp(\halo{\bx}|\both{\bf O})$. As $\halo{V}\fil$ is a real symmetric matrix, we can make use of the 
eigendecomposition $\halo{V}\fil = P\tp \Lambda P$ where $P$ is a matrix containing the eigenvectors of 
$\halo{V}\fil$ 
and $\Lambda = {\rm diag}(\lambda_1, \lambda_2, ..., \lambda_{2N})$, with $\lambda_i$ being an eigenvalue of 
$\halo{V}\fil$. Performing a change of basis into the eigenbasis, the Wigner function of the filtered state becomes
\beq\label{app_int1}
g(\check{\bz}; \ex{\hat{\bz}}\fil, PV\fil P\tp) = \int\dd\mu(\halo{\bz}) g(\halo{\bz};\halo{\bz}\fil,\Lambda) g(\check{\bz};\halo{\bz},P V\god P\tp)\,,
\eeq
where $\check{\bz} = P\bcx$, $\ex{\hat{\bz}} = P\xfil$, $\halo{\bz} = P\halo{\bx}$ and 
$\halo{\bz}\fil = P\halo{\bx}\fil$. Now, since $\Lambda$ is a diagonal matrix the Gaussian PDF can be factorized as 
$g(\halo{\bz};\halo{\bz}\fil,\Lambda) = \prod_{k = 1}^{2N} g(\halo{z}_k; \halo{z}_{{\rm F}, k}, \lambda_k)$ and we can 
consider the scenario where $\halo{V}\fil$ has at least one zero eigenvalue. 

Let us assume, without loss of generality, that $\lambda_k > 0\,\,\forall k\leq s$ and the remaining $2N - s$ 
eigenvalues are zero. Using the fact that $\lim_{\sigma\to 0} g(\check{x}; a, \sigma^2) = \delta(\check{x} - a)$, the 
Gaussian PDF becomes 
\beq
g(\halo{\bz};\halo{\bz}\fil,\Lambda) = \prod_{k = 1}^{s} g(\halo{z}_k; \halo{z}_{{\rm F}, k}, \lambda_k) \prod_{j = s+1}^{2N}
\delta(\halo{z}_j - \halo{z}_{{\rm F},k})\,.
\eeq
Computing the integral in \erf{app_int1}, we find that the transformed filtered mean and covariance are
\begin{align}
\ex{\hat{z}}\fil = [\halo{z}_{{\rm F},1}, ..., \halo{z}_{{\rm F},s}&, \ex{\hat z}_{{\rm T},s+1}, ..., \ex{\hat z}_{{\rm T},2N}]\tp\,,\\
PV\fil P\tp &= \Lambda + PV\god P\tp\,,\label{simp-hVf}
\end{align}
Thus, when $\halo{V}\fil$ has at least one zero eigenvalue, the corresponding components of the mean and 
covariance matrix are equal the same components of the transformed true mean and covariance. 

Moving on to quantum state smoothing, beginning with \erf{final_sm_int}, if we perform the same change of basis 
as we did for the filtered state, we obtain
\beq
\begin{split}
g(\check{\bz}; \ex{\hat{\bz}}\sm, PV\sm P\tp) = \int\dd\mu(\halo{\bz}) g&(\halo{\bz};\halo{\bz}\sm, P\halo{V}\sm P\tp) 
\\ \times &g(\check{\bz};\halo{\bz},P V\god P\tp)\,,
\end{split}
\eeq
where $\ex{\hat{\bz}}\sm = P\xsm$ and $\halo{\bz}\sm = P\halo{\bx}\sm$.
However, unlike in the filtering case, $P\halo{V}\sm P\tp$ is not necessarily a diagonal matrix and thus the 
Gaussian cannot be completely factorized. However, we can still expect the Gaussian to factorize as
\beq\label{app_int2}
g(\halo{\bz};\halo{\bz}\sm, P\halo{V}\sm P\tp) = g(\halo{\bz}'; \halo{\bz}\sm', \halo{V}\sm') \prod_{j = s+1}^{2N} \delta(\halo{z}_j - \halo{z}_{{\rm S}, j})\,,
\eeq
because conditioning the estimate of $\ex{\hat{\bz}}\god$ on more information cannot make those components of 
the probability distribution any more certain than a delta function. 
Here $\halo{\bz}'$, $\halo{\bz}\sm'$ are the first $s$ components of $\halo{\bz}$ and $\halo{\bz}\sm$, respectively, 
and $\halo{V}\sm'$ is the first $s\times s$ block of $P\halo{V}\sm P\tp$. Computing the integral in 
\erf{app_int2} gives
\begin{align}
\ex{\hat{\bz}}\sm = [\halo{\bz}'\sm&, \ex{\hat z}_{{\rm T},s+1}, ..., \ex{\hat z}_{{\rm T},2N}]\tp\,,\label{app_sm_mean}\\
PV\sm P\tp &= \left[\begin{matrix}
\halo{V}\sm' & 0\\
0 & 0
\end{matrix}
\right] + PV\god P\tp\,.\label{app_sm_V}
\end{align}
We see that, like the filtered state, the components of the smoothed mean that have an eigenvalue of zero for the 
$\hV\fil(t)$ are equal corresponding components of the true mean and similarly for 
the smoothed covariance matrix. The remaining components of the mean and covariance are computed using the 
remaining elements of the filtered mean and covariance. This now gives use a method to compute the smoothed 
quantum state when $\hV\fil\inv$ does not exist at a particular time $t$. I will comment that in the event that this 
occurs, the MFP form in Ref.~\cite{LCW19} will be more useful because when the inverse does not exist at time 
$t$, it will only affect the smoothed mean and covariance at $t$ which can easily be correct. 

As an aside, this analysis highlights an interesting property of the smoothed quantum state. In the event that all 
the eigenvalues of the haloed filtered covariance are zero, using \erfs{app_sm_mean}{app_sm_V} the smoothed 
mean and covariance will be equal to the true mean and true covariance, respectively. While this in itself is not 
particularly interesting, it becomes interesting when we consider the initial conditions for the smoothed state. As 
we have assumed throughout this paper we have $V\fil(t_0) = V\god(t_0) = V_0$, i.e., $\halo{V}\fil(t_0) = 0$. Thus, 
we have that initially the smoothed quantum state must have mean $\xsm(t_0) = \xgod(t_0) = \ex{\hbx}_0$ and 
covariance $V\sm(t_0) = V\god(t_0) = V_0$, meaning that, in this case, the smoothed quantum state will always 
start in the same state as both the filtered and true quantum state. Moreover, this fact holds irrespective of 
whether the true state is pure. As pointed out earlier, we can see this occur in Fig.~\ref{Fig-1}, where we also see 
that the classically smoothed state does not coincide with the filtered and true means initially. This is because, in 
the classical case, it is always assumed that the classical true state (the state of maximal knowledge) is a delta 
function causing the condition to be violated.  Note, this does not mean that the quantum smoothed mean can 
be computed forward in time, just that it is constrained at both $t_0$ and $T$. 

While the MFP form of the quantum state smoothing equations might be more useful in general, there is a special 
case when $P$ is time independent over an interval $[\tau_1, \tau_2]$ with $\lambda_k = 0$ for 
$k>s$. In this case, the Moore-Penrose pseudo-inverse can be used in \erfs{dxs}{dVs} instead of the usual matrix 
inverse. Over this interval, 
following the reasoning prior to this, $\ex{\hat\bz}\sm$ and $\hV\sm$ will be of the forms \erf{app_sm_mean} and 
(\ref{app_sm_V}), respectively, over the interval. I will show that taking the pseudo-inverse causes the mean and 
haloed covariance matrix (as this is simpler to show analytically) allows the relevant components to evolve as the 
corresponding components of the true state would, while the remaining components evolve in a similar manner to 
a system where the inverse exists. 

Beginning with the mean, the stochastic differential equation for the 
transformed mean is
\beq
\dd\zsm = \bar{A}^{*}\zsm\dd t + \bar{D}^{*} \Lambda^{+}(\zsm -\ex{\hat\bz}\fil)\dd t + P\K\ob^+[V\god]\by\ob\dd t\,,
\eeq
where I have used $P\tp P = I_{2N}$, an asterisked matrix denotes a transformation by $P$, i.e. 
$F^* = PFP\tp$ and $\Lambda^{+}$ is the pseudo-inverse of $\Lambda$. For a diagonal matrix, the pseudo-inverse is simple to compute by inverting the non-zero elements and leaving the remaining elements unchanged. 
Looking at the $k$th component of the mean we have
\beq\label{element_sm}
\begin{split}
\dd\ex{\hat{z}_k}\sm = \sum_\ell \bar{A}^{*}_{k,\ell}&\ex{\hat{z}_\ell}\sm\dd t + (\bar{D}\Lambda)_{k,\ell} (\ex{\hat{z}_\ell}\sm \\- &\ex{\hat{z}_\ell}\fil)\dd t + (P\K\ob^+[V\god])_{k,\ell}y_{{\rm o},\ell}\dd t\,,
\end{split}
\eeq
where for comparison, the evolution of the $k$th component of the transformed true mean is
\beq\label{element_true}
\begin{split}
\dd\ex{\hat{z}_k}\god = \sum_\ell \bar{A}^{*}_{k,\ell}\ex{\hat{z}_\ell}\god\dd t +& (P\K\ob^+[V\god])_{k,\ell}y_{{\rm o},\ell}\dd t \\+& (P\K\un^+[V\god])_{k,\ell}\dd w_{{\rm u},\ell}\,.
\end{split}
\eeq
When $k > s$, since the evolution of the smoothed mean must be the same as the evolution of the true mean over the time interval, as $\ex{\hat z_k}\sm(\tau_2) = \ex{\hat z_k}\god(\tau_2)$, it must be the case that the matrices must have the following block forms:
\beq\label{block_form}
\bar{A}^{*} = \left[\begin{matrix}
\bar{A}^*_{00} & \bar{A}^*_{01} \\
0 & \bar{A}^*_{11}
\end{matrix}\right]\,,\qquad \bar{D}^{*} = \left[\begin{matrix}
\bar{D}^*_{00} & 0\\
0 & 0
\end{matrix}\right]\,,
\eeq
where the blocks are divided so that the diagonal matrices have dimensions $s \times s$ and 
$(2N - s) \times (2N - s)$. 

We can understand why this must be the case because if $\bar{A}^*$ was of 
another form the first term in \erf{element_sm} would cause the evolution of the $k$th component to be influenced 
by components other than the true mean and hence would cause the $k$th components to deviate from the true 
mean. A similar reasoning follows for the form of $\bar{D}^*$ to eliminate the second term. Note, since the 
lower half of $\bar{D}^*$ is zero, we see, using $\bar{D} = \K\un^+[V\god]\K\un^+[V\god]\tp$, that 
$(\K\un^+[V\god])_{k,\ell} = 0$ for all $\ell$ when $k > s$. This will eliminate the final term in \erf{element_true} and 
the $k$th components will evolve identically. 

All that remains is to show that $\bar{A}^*$ and $\bar{D}^*$ are of the forms in \erf{block_form}. 
Consider the differential equation for the haloed filtered covariance \cite{LCW19,LCW-PRA21},
\beq
\ddt{\halo{V}\fil} = \bar{A}\hV\fil + \hV\fil\bar{A}\tp + \bar{D} - \hV\fil C\ob\tp C\ob\hV\fil\,.
\eeq
Using $\hV\fil = P\tp\Lambda(t) P$ we obtain
\beq
\ddt{\Lambda} = \bar{A}^*\Lambda + \Lambda\bar{A}^{*\top} + \bar{D}^* - \Lambda P\tp C\ob\tp C\ob P\Lambda\,,
\eeq
and for the $k,\ell$th element we have
\beq\label{element_lambda}
\ddt{\lambda_k}\delta_{k,\ell} = \lambda_\ell\bar{A}^*_{k,\ell} + \lambda_k \bar{A}^{*\top}_{k,\ell} + \bar{D}^*_{k,\ell} - \lambda_k\lambda_\ell (PC\ob\tp C\ob P\tp)_{k,\ell}\,,
\eeq
where Einstein's summation convention is {\em not} being used. Looking at $k, \ell > s$ and 
remembering that the zero eigenvalues do not change over the interval, we have
$\bar{D}^*_{k,\ell} = 0$.
By looking at the $k,k$th element of $\bar{D}^*$ with $k > s$, we have $\bar{D}^* = \sum_\ell (P\K\un^+[V\god])_{k,\ell}^2 = 0$ and thus $(P\K\un^+[V\god])_{k,\ell} = 0$ for all $\ell$, showing that $\bar{D}^*$ is of 
the form in \erf{block_form}. Now, returning to \erf{element_lambda}, if we consider the case where $k > s$ and $\ell < s + 1$, we have $\lambda_\ell \bar{A}'_{k,\ell} + \bar{D}^*_{k,\ell} = 0$. As we have just shown, in this regime 
$\bar{D}^*_{k,\ell} = 0$, and we know $\lambda_\ell \ne 0$, resulting in $\bar{A}^*_{k,\ell} = 0$. Thus $\bar{A}^*$ is also of the form in \erf{block_form} and completes the proof that the components of the transformed smoothed mean with $k > s$ are equal to the true mean during the interval.
Moving onto the haloed covariance matrix, consider the transformed differential equation for the haloed smoothed 
is 
\beq
\ddt{\hV\sm^*} = (\bar{A}^* +\bar{D}^*\Lambda^+)\hV\sm^* + \hV\sm^* (\bar{A}^* 
+\bar{D}^*\Lambda^+)\tp - \bar{D}^*\,.
\eeq
It is easy to show using the \erf{block_form} that, given the haloed covariance is of the form 
\beq
\hV\sm^* = \left[\begin{matrix}
(\hV\sm^*)_{00} & 0\\
0 & 0
\end{matrix}\right]\,,
\eeq
as is the case at $\tau_2$, then the covariance will remain in that form.

%\bibliography{Thesis-Bib.bib}

\begin{thebibliography}{33}%
\makeatletter
\providecommand \@ifxundefined [1]{%
 \@ifx{#1\undefined}
}%
\providecommand \@ifnum [1]{%
 \ifnum #1\expandafter \@firstoftwo
 \else \expandafter \@secondoftwo
 \fi
}%
\providecommand \@ifx [1]{%
 \ifx #1\expandafter \@firstoftwo
 \else \expandafter \@secondoftwo
 \fi
}%
\providecommand \natexlab [1]{#1}%
\providecommand \enquote  [1]{``#1''}%
\providecommand \bibnamefont  [1]{#1}%
\providecommand \bibfnamefont [1]{#1}%
\providecommand \citenamefont [1]{#1}%
\providecommand \href@noop [0]{\@secondoftwo}%
\providecommand \href [0]{\begingroup \@sanitize@url \@href}%
\providecommand \@href[1]{\@@startlink{#1}\@@href}%
\providecommand \@@href[1]{\endgroup#1\@@endlink}%
\providecommand \@sanitize@url [0]{\catcode `\\12\catcode `\$12\catcode
  `\&12\catcode `\#12\catcode `\^12\catcode `\_12\catcode `\%12\relax}%
\providecommand \@@startlink[1]{}%
\providecommand \@@endlink[0]{}%
\providecommand \url  [0]{\begingroup\@sanitize@url \@url }%
\providecommand \@url [1]{\endgroup\@href {#1}{\urlprefix }}%
\providecommand \urlprefix  [0]{URL }%
\providecommand \Eprint [0]{\href }%
\providecommand \doibase [0]{http://dx.doi.org/}%
\providecommand \selectlanguage [0]{\@gobble}%
\providecommand \bibinfo  [0]{\@secondoftwo}%
\providecommand \bibfield  [0]{\@secondoftwo}%
\providecommand \translation [1]{[#1]}%
\providecommand \BibitemOpen [0]{}%
\providecommand \bibitemStop [0]{}%
\providecommand \bibitemNoStop [0]{.\EOS\space}%
\providecommand \EOS [0]{\spacefactor3000\relax}%
\providecommand \BibitemShut  [1]{\csname bibitem#1\endcsname}%
\let\auto@bib@innerbib\@empty
%</preamble>
\bibitem [{\citenamefont {Kalman}\ and\ \citenamefont {Bucy}(1961)}]{KalBuc61}%
  \BibitemOpen
  \bibfield  {author} {\bibinfo {author} {\bibfnamefont {R.~E.}\ \bibnamefont
  {Kalman}}\ and\ \bibinfo {author} {\bibfnamefont {R.~S.}\ \bibnamefont
  {Bucy}},\ }\href@noop {} {\bibfield  {journal} {\bibinfo  {journal} {Journal
  of basic engineering}\ }\textbf {\bibinfo {volume} {83}},\ \bibinfo {pages}
  {95} (\bibinfo {year} {1961})}\BibitemShut {NoStop}%
\bibitem [{\citenamefont {Rauch}(1963)}]{Rauch63}%
  \BibitemOpen
  \bibfield  {author} {\bibinfo {author} {\bibfnamefont {H.}~\bibnamefont
  {Rauch}},\ }\href {\doibase 10.1109/TAC.1963.1105600} {\bibfield  {journal}
  {\bibinfo  {journal} {IEEE Transactions on Automatic Control}\ }\textbf
  {\bibinfo {volume} {8}},\ \bibinfo {pages} {371} (\bibinfo {year}
  {1963})}\BibitemShut {NoStop}%
\bibitem [{\citenamefont {Rauch}\ \emph {et~al.}(1965)\citenamefont {Rauch},
  \citenamefont {Tung},\ and\ \citenamefont {Striebel}}]{RTS65}%
  \BibitemOpen
  \bibfield  {author} {\bibinfo {author} {\bibfnamefont {H.~E.}\ \bibnamefont
  {Rauch}}, \bibinfo {author} {\bibfnamefont {F.}~\bibnamefont {Tung}}, \ and\
  \bibinfo {author} {\bibfnamefont {C.~T.}\ \bibnamefont {Striebel}},\ }\href
  {\doibase 10.2514/3.3166} {\bibfield  {journal} {\bibinfo  {journal} {AIAA
  journal}\ }\textbf {\bibinfo {volume} {3}},\ \bibinfo {pages} {1445}
  (\bibinfo {year} {1965})}\BibitemShut {NoStop}%
\bibitem [{\citenamefont {Mayne}(1966)}]{Mayne66}%
  \BibitemOpen
  \bibfield  {author} {\bibinfo {author} {\bibfnamefont {D.~Q.}\ \bibnamefont
  {Mayne}},\ }\href {\doibase 10.1016/0005-1098(66)90019-7} {\bibfield
  {journal} {\bibinfo  {journal} {Automatica}\ }\textbf {\bibinfo {volume}
  {4}},\ \bibinfo {pages} {73} (\bibinfo {year} {1966})}\BibitemShut {NoStop}%
\bibitem [{\citenamefont {Fraser}(1967)}]{Fraser67}%
  \BibitemOpen
  \bibfield  {author} {\bibinfo {author} {\bibfnamefont {D.~C.}\ \bibnamefont
  {Fraser}},\ }\emph {\bibinfo {title} {A new technique for the optimal
  smoothing of data}},\ \href@noop {} {Ph.D. thesis},\ \bibinfo  {school}
  {Massachusetts Institute of Technology} (\bibinfo {year} {1967})\BibitemShut
  {NoStop}%
\bibitem [{\citenamefont {Fraser}\ and\ \citenamefont
  {Potter}(1969)}]{FraPot69}%
  \BibitemOpen
  \bibfield  {author} {\bibinfo {author} {\bibfnamefont {D.}~\bibnamefont
  {Fraser}}\ and\ \bibinfo {author} {\bibfnamefont {J.}~\bibnamefont
  {Potter}},\ }\href {\doibase 10.1109/TAC.1969.1099196} {\bibfield  {journal}
  {\bibinfo  {journal} {IEEE Transactions on automatic control}\ }\textbf
  {\bibinfo {volume} {14}},\ \bibinfo {pages} {387} (\bibinfo {year}
  {1969})}\BibitemShut {NoStop}%
\bibitem [{\citenamefont {Weinert}(2001)}]{Weinert01}%
  \BibitemOpen
  \bibfield  {author} {\bibinfo {author} {\bibfnamefont {H.~L.}\ \bibnamefont
  {Weinert}},\ }\href@noop {} {\emph {\bibinfo {title} {Fixed Interval
  Smoothing for State Space Models}}}\ (\bibinfo  {publisher} {Kluwer
  Academic},\ \bibinfo {address} {New York},\ \bibinfo {year}
  {2001})\BibitemShut {NoStop}%
\bibitem [{\citenamefont {Haykin}(2001)}]{Hay01}%
  \BibitemOpen
  \bibfield  {author} {\bibinfo {author} {\bibfnamefont {S.}~\bibnamefont
  {Haykin}},\ }\href@noop {} {\emph {\bibinfo {title} {Kalman Filtering and
  Neural Networks}}}\ (\bibinfo  {publisher} {Wiley},\ \bibinfo {address} {New
  York},\ \bibinfo {year} {2001})\BibitemShut {NoStop}%
\bibitem [{\citenamefont {Trees}\ and\ \citenamefont
  {Bell}(2013)}]{vanTrees13}%
  \BibitemOpen
  \bibfield  {author} {\bibinfo {author} {\bibfnamefont {H.~L.~V.}\
  \bibnamefont {Trees}}\ and\ \bibinfo {author} {\bibfnamefont {K.~L.}\
  \bibnamefont {Bell}},\ }\href@noop {} {\emph {\bibinfo {title} {Detection,
  Estimation, and Modulation Theory, Part I: Detection, Estimation, and
  Filtering Theory}}},\ \bibinfo {edition} {2nd}\ ed.\ (\bibinfo  {publisher}
  {John Wiley and Sons},\ \bibinfo {address} {New York},\ \bibinfo {year}
  {2013})\BibitemShut {NoStop}%
\bibitem [{\citenamefont {Brown}\ and\ \citenamefont {Hwang}(2012)}]{BroHwa12}%
  \BibitemOpen
  \bibfield  {author} {\bibinfo {author} {\bibfnamefont {R.~G.}\ \bibnamefont
  {Brown}}\ and\ \bibinfo {author} {\bibfnamefont {P.~Y.~C.}\ \bibnamefont
  {Hwang}},\ }\href@noop {} {\emph {\bibinfo {title} {Introduction to Random
  Signals and Applied Kalman Filtering}}},\ \bibinfo {edition} {4th}\ ed.\
  (\bibinfo  {publisher} {Wiley},\ \bibinfo {address} {New York},\ \bibinfo
  {year} {2012})\BibitemShut {NoStop}%
\bibitem [{\citenamefont {Einicke}(2012)}]{Einicke12}%
  \BibitemOpen
  \bibfield  {author} {\bibinfo {author} {\bibfnamefont {G.~A.}\ \bibnamefont
  {Einicke}},\ }\href@noop {} {\emph {\bibinfo {title} {Smoothing, filtering
  and prediction: Estimating the past, present and future}}}\ (\bibinfo
  {publisher} {InTech Rijeka},\ \bibinfo {year} {2012})\BibitemShut {NoStop}%
\bibitem [{\citenamefont {Friedland}(2012)}]{Fri12}%
  \BibitemOpen
  \bibfield  {author} {\bibinfo {author} {\bibfnamefont {B.}~\bibnamefont
  {Friedland}},\ }\href@noop {} {\emph {\bibinfo {title} {Control system
  design: an introduction to state-space methods}}}\ (\bibinfo  {publisher}
  {Courier Corporation},\ \bibinfo {year} {2012})\BibitemShut {NoStop}%
\bibitem [{\citenamefont {S{\"a}rkk{\"a}}(2013)}]{Sarkka13}%
  \BibitemOpen
  \bibfield  {author} {\bibinfo {author} {\bibfnamefont {S.}~\bibnamefont
  {S{\"a}rkk{\"a}}},\ }\href@noop {} {\emph {\bibinfo {title} {Bayesian
  filtering and smoothing}}},\ Vol.~\bibinfo {volume} {3}\ (\bibinfo
  {publisher} {Cambridge University Press},\ \bibinfo {year}
  {2013})\BibitemShut {NoStop}%
\bibitem [{\citenamefont {Belavkin}(1987)}]{Bel87}%
  \BibitemOpen
  \bibfield  {author} {\bibinfo {author} {\bibfnamefont {V.~P.}\ \bibnamefont
  {Belavkin}},\ }\href@noop {} {\emph {\bibinfo {title} {Information,
  complexity and control in quantum physics}}},\ edited by\ \bibinfo {editor}
  {\bibfnamefont {A.}~\bibnamefont {Blaqu\'{i}ere}}, \bibinfo {editor}
  {\bibfnamefont {S.}~\bibnamefont {Dinar}}, \ and\ \bibinfo {editor}
  {\bibfnamefont {G.}~\bibnamefont {Lochak}}\ (\bibinfo  {publisher}
  {Springer},\ \bibinfo {address} {New York},\ \bibinfo {year}
  {1987})\BibitemShut {NoStop}%
\bibitem [{\citenamefont {Belavkin}(1992)}]{Bel92}%
  \BibitemOpen
  \bibfield  {author} {\bibinfo {author} {\bibfnamefont {V.~P.}\ \bibnamefont
  {Belavkin}},\ }\href {\doibase 10.1007/BF02097018} {\bibfield  {journal}
  {\bibinfo  {journal} {Communications in Mathematical Physics}\ }\textbf
  {\bibinfo {volume} {146}},\ \bibinfo {pages} {611} (\bibinfo {year}
  {1992})}\BibitemShut {NoStop}%
\bibitem [{\citenamefont {Belavkin}(1999)}]{Bel99}%
  \BibitemOpen
  \bibfield  {author} {\bibinfo {author} {\bibfnamefont {V.~P.}\ \bibnamefont
  {Belavkin}},\ }\href {\doibase doi:10.1016/S0034-4877(00)86386-7} {\bibfield
  {journal} {\bibinfo  {journal} {Rep. Math. Phys.}\ }\textbf {\bibinfo
  {volume} {43}},\ \bibinfo {pages} {A405} (\bibinfo {year}
  {1999})}\BibitemShut {NoStop}%
\bibitem [{\citenamefont {Doherty}\ and\ \citenamefont
  {Jacobs}(1999)}]{DohJac99}%
  \BibitemOpen
  \bibfield  {author} {\bibinfo {author} {\bibfnamefont {A.~C.}\ \bibnamefont
  {Doherty}}\ and\ \bibinfo {author} {\bibfnamefont {K.}~\bibnamefont
  {Jacobs}},\ }\href {\doibase 10.1103/PhysRevA.60.2700} {\bibfield  {journal}
  {\bibinfo  {journal} {Phys. Rev. A}\ }\textbf {\bibinfo {volume} {60}},\
  \bibinfo {pages} {2700} (\bibinfo {year} {1999})}\BibitemShut {NoStop}%
\bibitem [{\citenamefont {Wiseman}\ and\ \citenamefont
  {Doherty}(2005)}]{WisDoh05}%
  \BibitemOpen
  \bibfield  {author} {\bibinfo {author} {\bibfnamefont {H.~M.}\ \bibnamefont
  {Wiseman}}\ and\ \bibinfo {author} {\bibfnamefont {A.~C.}\ \bibnamefont
  {Doherty}},\ }\href {\doibase 10.1103/PhysRevLett.94.070405} {\bibfield
  {journal} {\bibinfo  {journal} {Phys. Rev. Lett.}\ }\textbf {\bibinfo
  {volume} {94}},\ \bibinfo {pages} {070405} (\bibinfo {year}
  {2005})}\BibitemShut {NoStop}%
\bibitem [{\citenamefont {Wiseman}\ and\ \citenamefont
  {Milburn}(2010)}]{WisMil10}%
  \BibitemOpen
  \bibfield  {author} {\bibinfo {author} {\bibfnamefont {H.~M.}\ \bibnamefont
  {Wiseman}}\ and\ \bibinfo {author} {\bibfnamefont {G.~J.}\ \bibnamefont
  {Milburn}},\ }\href@noop {} {\emph {\bibinfo {title} {Quantum Measurement and
  Control}}}\ (\bibinfo  {publisher} {Cambridge University Press},\ \bibinfo
  {address} {Cambridge, England},\ \bibinfo {year} {2010})\BibitemShut
  {NoStop}%
\bibitem [{\citenamefont {Tsang}(2009)}]{Tsang-PRA09}%
  \BibitemOpen
  \bibfield  {author} {\bibinfo {author} {\bibfnamefont {M.}~\bibnamefont
  {Tsang}},\ }\href {\doibase 10.1103/PhysRevA.80.033840} {\bibfield  {journal}
  {\bibinfo  {journal} {Phys. Rev. A}\ }\textbf {\bibinfo {volume} {80}},\
  \bibinfo {pages} {033840} (\bibinfo {year} {2009})}\BibitemShut {NoStop}%
\bibitem [{\citenamefont {Gammelmark}\ \emph {et~al.}(2013)\citenamefont
  {Gammelmark}, \citenamefont {Julsgaard},\ and\ \citenamefont
  {M\o{}lmer}}]{GJM13}%
  \BibitemOpen
  \bibfield  {author} {\bibinfo {author} {\bibfnamefont {S.}~\bibnamefont
  {Gammelmark}}, \bibinfo {author} {\bibfnamefont {B.}~\bibnamefont
  {Julsgaard}}, \ and\ \bibinfo {author} {\bibfnamefont {K.}~\bibnamefont
  {M\o{}lmer}},\ }\href {\doibase PhysRevLett.111.160401} {\bibfield  {journal}
  {\bibinfo  {journal} {Phys. Rev. Lett.}\ }\textbf {\bibinfo {volume} {{\bf
  111}}},\ \bibinfo {pages} {160401} (\bibinfo {year} {2013})}\BibitemShut
  {NoStop}%
\bibitem [{\citenamefont {Guevara}\ and\ \citenamefont
  {Wiseman}(2015)}]{GueWis15}%
  \BibitemOpen
  \bibfield  {author} {\bibinfo {author} {\bibfnamefont {I.}~\bibnamefont
  {Guevara}}\ and\ \bibinfo {author} {\bibfnamefont {H.}~\bibnamefont
  {Wiseman}},\ }\href {\doibase 10.1103/PhysRevLett.115.180407} {\bibfield
  {journal} {\bibinfo  {journal} {Phys. Rev. Lett.}\ }\textbf {\bibinfo
  {volume} {115}},\ \bibinfo {pages} {180407} (\bibinfo {year}
  {2015})}\BibitemShut {NoStop}%
\bibitem [{\citenamefont {Ohki}(2015)}]{Ohki15}%
  \BibitemOpen
  \bibfield  {author} {\bibinfo {author} {\bibfnamefont {K.}~\bibnamefont
  {Ohki}},\ }in\ \href {\doibase 10.1109/CDC.2015.7402898} {\emph {\bibinfo
  {booktitle} {2015 54th IEEE Conference on Decision and Control (CDC)}}}\
  (\bibinfo {year} {2015})\ pp.\ \bibinfo {pages} {4350--4355}\BibitemShut
  {NoStop}%
\bibitem [{\citenamefont {Laverick}\ \emph
  {et~al.}(2021{\natexlab{a}})\citenamefont {Laverick}, \citenamefont
  {Chantasri},\ and\ \citenamefont {Wiseman}}]{LCW-QS20}%
  \BibitemOpen
  \bibfield  {author} {\bibinfo {author} {\bibfnamefont {K.~T.}\ \bibnamefont
  {Laverick}}, \bibinfo {author} {\bibfnamefont {A.}~\bibnamefont {Chantasri}},
  \ and\ \bibinfo {author} {\bibfnamefont {H.~M.}\ \bibnamefont {Wiseman}},\
  }\href {\doibase 10.1007/s40509-020-00225-7} {\bibfield  {journal} {\bibinfo
  {journal} {Quantum Stud.: Math. Found.}\ }\textbf {\bibinfo {volume} {8}},\
  \bibinfo {pages} {37} (\bibinfo {year} {2021}{\natexlab{a}})}\BibitemShut
  {NoStop}%
\bibitem [{\citenamefont {Chantasri}\ \emph {et~al.}(2019)\citenamefont
  {Chantasri}, \citenamefont {Guevara},\ and\ \citenamefont {Wiseman}}]{CGW19}%
  \BibitemOpen
  \bibfield  {author} {\bibinfo {author} {\bibfnamefont {A.}~\bibnamefont
  {Chantasri}}, \bibinfo {author} {\bibfnamefont {I.}~\bibnamefont {Guevara}},
  \ and\ \bibinfo {author} {\bibfnamefont {H.~M.}\ \bibnamefont {Wiseman}},\
  }\href {\doibase 10.1088/1367-2630/ab396e} {\bibfield  {journal} {\bibinfo
  {journal} {New Journal of Physics}\ }\textbf {\bibinfo {volume} {21}},\
  \bibinfo {pages} {083039} (\bibinfo {year} {2019})}\BibitemShut {NoStop}%
\bibitem [{\citenamefont {Chantasri}\ \emph {et~al.}(2021)\citenamefont
  {Chantasri}, \citenamefont {Guevara}, \citenamefont {Laverick},\ and\
  \citenamefont {Wiseman}}]{CGLW21}%
  \BibitemOpen
  \bibfield  {author} {\bibinfo {author} {\bibfnamefont {A.}~\bibnamefont
  {Chantasri}}, \bibinfo {author} {\bibfnamefont {I.}~\bibnamefont {Guevara}},
  \bibinfo {author} {\bibfnamefont {K.~T.}\ \bibnamefont {Laverick}}, \ and\
  \bibinfo {author} {\bibfnamefont {H.~M.}\ \bibnamefont {Wiseman}},\
  }\href@noop {} {\  (\bibinfo {year} {2021})},\ \Eprint
  {http://arxiv.org/abs/2104.02911} {arXiv:2104.02911 [quant-ph]} \BibitemShut
  {NoStop}%
\bibitem [{\citenamefont {Laverick}\ \emph {et~al.}(2019)\citenamefont
  {Laverick}, \citenamefont {Chantasri},\ and\ \citenamefont
  {Wiseman}}]{LCW19}%
  \BibitemOpen
  \bibfield  {author} {\bibinfo {author} {\bibfnamefont {K.~T.}\ \bibnamefont
  {Laverick}}, \bibinfo {author} {\bibfnamefont {A.}~\bibnamefont {Chantasri}},
  \ and\ \bibinfo {author} {\bibfnamefont {H.~M.}\ \bibnamefont {Wiseman}},\
  }\href {\doibase 10.1103/PhysRevLett.122.190402} {\bibfield  {journal}
  {\bibinfo  {journal} {Phys. Rev. Lett.}\ }\textbf {\bibinfo {volume} {122}},\
  \bibinfo {pages} {190402} (\bibinfo {year} {2019})}\BibitemShut {NoStop}%
\bibitem [{\citenamefont {Laverick}\ \emph
  {et~al.}(2021{\natexlab{b}})\citenamefont {Laverick}, \citenamefont
  {Chantasri},\ and\ \citenamefont {Wiseman}}]{LCW-PRA21}%
  \BibitemOpen
  \bibfield  {author} {\bibinfo {author} {\bibfnamefont {K.~T.}\ \bibnamefont
  {Laverick}}, \bibinfo {author} {\bibfnamefont {A.}~\bibnamefont {Chantasri}},
  \ and\ \bibinfo {author} {\bibfnamefont {H.~M.}\ \bibnamefont {Wiseman}},\
  }\href {\doibase 10.1103/PhysRevA.103.012213} {\bibfield  {journal} {\bibinfo
   {journal} {Phys. Rev. A}\ }\textbf {\bibinfo {volume} {103}},\ \bibinfo
  {pages} {012213} (\bibinfo {year} {2021}{\natexlab{b}})}\BibitemShut
  {NoStop}%
\bibitem [{\citenamefont {Kailath}\ and\ \citenamefont
  {Frost}(1968)}]{KaiFro68}%
  \BibitemOpen
  \bibfield  {author} {\bibinfo {author} {\bibfnamefont {T.}~\bibnamefont
  {Kailath}}\ and\ \bibinfo {author} {\bibfnamefont {P.}~\bibnamefont
  {Frost}},\ }\href@noop {} {\bibfield  {journal} {\bibinfo  {journal} {IEEE
  Transactions on Automatic Control}\ }\textbf {\bibinfo {volume} {13}},\
  \bibinfo {pages} {655} (\bibinfo {year} {1968})}\BibitemShut {NoStop}%
\bibitem [{\citenamefont {Kailath}(1970)}]{Kai70}%
  \BibitemOpen
  \bibfield  {author} {\bibinfo {author} {\bibfnamefont {T.}~\bibnamefont
  {Kailath}},\ }\href@noop {} {\bibfield  {journal} {\bibinfo  {journal}
  {Proceedings of the IEEE}\ }\textbf {\bibinfo {volume} {58}},\ \bibinfo
  {pages} {680} (\bibinfo {year} {1970})}\BibitemShut {NoStop}%
\bibitem [{\citenamefont {Kailath}(1973)}]{Kai73}%
  \BibitemOpen
  \bibfield  {author} {\bibinfo {author} {\bibfnamefont {T.}~\bibnamefont
  {Kailath}},\ }\href@noop {} {\bibfield  {journal} {\bibinfo  {journal} {IEEE
  transactions on Information Theory}\ }\textbf {\bibinfo {volume} {19}},\
  \bibinfo {pages} {750} (\bibinfo {year} {1973})}\BibitemShut {NoStop}%
\bibitem [{\citenamefont {Badawi}\ \emph {et~al.}(1979)\citenamefont {Badawi},
  \citenamefont {Lindquist},\ and\ \citenamefont {Pavon}}]{BLP79}%
  \BibitemOpen
  \bibfield  {author} {\bibinfo {author} {\bibfnamefont {F.}~\bibnamefont
  {Badawi}}, \bibinfo {author} {\bibfnamefont {A.}~\bibnamefont {Lindquist}}, \
  and\ \bibinfo {author} {\bibfnamefont {M.}~\bibnamefont {Pavon}},\
  }\href@noop {} {\bibfield  {journal} {\bibinfo  {journal} {IEEE Transactions
  on Automatic Control}\ }\textbf {\bibinfo {volume} {24}},\ \bibinfo {pages}
  {878} (\bibinfo {year} {1979})}\BibitemShut {NoStop}%
\bibitem [{\citenamefont {Chia}\ and\ \citenamefont
  {Wiseman}(2011)}]{ChiWis11}%
  \BibitemOpen
  \bibfield  {author} {\bibinfo {author} {\bibfnamefont {A.}~\bibnamefont
  {Chia}}\ and\ \bibinfo {author} {\bibfnamefont {H.~M.}\ \bibnamefont
  {Wiseman}},\ }\href {\doibase 10.1103/PhysRevA.84.012119} {\bibfield
  {journal} {\bibinfo  {journal} {Physical Review A}\ }\textbf {\bibinfo
  {volume} {84}},\ \bibinfo {pages} {012119} (\bibinfo {year}
  {2011})}\BibitemShut {NoStop}%
\end{thebibliography}
%merlin.mbs apsrev4-1.bst 2010-07-25 4.21a (PWD, AO, DPC) hacked
%Control: key (0)
%Control: author (8) initials jnrlst
%Control: editor formatted (1) identically to author
%Control: production of article title (-1) disabled
%Control: page (0) single
%Control: year (1) truncated
%Control: production of eprint (0) enabled
%

\end{document}